\documentclass{aa} 
\usepackage{graphicx} 
\usepackage{natbib}
\bibpunct{(}{)}{;}{a}{}{,}
\usepackage{lscape} \usepackage{longtable} \usepackage{ulem}
\def\new#1 {{\bf #1} } \def\cut#1 {\sout{#1}}
\bibpunct{(}{)}{;}{a}{}{,}
\usepackage{subfigure}
\usepackage{txfonts}
\begin{document}
\title{Multi-line (sub)millimetre observations of the high-mass proto
cluster IRAS 05358+3543} \author{S. Leurini\inst{1,2},
H. Beuther\inst{3}, P.  Schilke\inst{1}, F. Wyrowski\inst{1},
Q. Zhang\inst{4} \& K.M. Menten\inst{1}}
\institute{Max-Planck-Institut f\"ur Radioastronomie, Auf dem H\"ugel
69, D-53121, Bonn  \and ESO, Karl-Schwarzschild Strasse 2, D-85748,
Garching-bei-M\"unchen \\ \email{sleurini@eso.org} \and Max-Planck-Institut f\"ur Astronomie,
K\"onigstuhl 17, D-69117, Heidelberg \and Harvard-Smithsonian Center
for Astrophysics, 60 Garden Street, Cambridge, MA 02138, USA}
\offprints{S. Leurini} 
\date{\today} 
\abstract {Since most high- and intermediate-mass 
protostars are at great distance and form in clusters, 
high linear resolution observations are
needed to investigate their physical properties.}
{To study the gas in the innermost region around the protostars in the
proto-cluster IRAS\,05358+3543, we observed the source in several
transitions of methanol and other molecular species with
the Plateau de Bure Interferometer and the Submillimeter Array,
reaching a linear resolution of 1100~AU.} {We determine the kinetic temperature of the gas around
the protostars through an LVG and LTE analysis of their molecular
emission; the column densities of 
CH$_3$OH, CH$_3$CN and SO$_2$ are also derived. Constrains on the density of the gas are estimated for two of the protostellar cores.} {We
find that the dust condensations are in various evolutionary stages. The
powerhouse of the cluster, mm1a, harbours a hot core with
$T\sim 220~(75<T<330)$~K. A double-peaked profile is detected in several
transitions toward mm1a, and we found a velocity gradient along a linear
structure which could be perpendicular to one of the outflows from the
vicinity of mm1a. Since the size of the double-peaked emission is less
than 1100~AU, we suggest that mm1a might host a massive circumstellar
disk. The other sources are in earlier stages of star
formation. The least active source, mm3, could be a starless massive
core, since it is cold ($T<20$~K), with a large reservoir of accreting
material ($M\sim 19~M_\odot$), but no molecular emission peaks on it.} {}
 \titlerunning{(Sub)mm multi-line observations of
IRAS\,05358+3543} \authorrunning{Leurini et al.} \maketitle

\section{Introduction}

The last decade has seen  significant progress in the understanding
of how high-mass stars form. Large samples of massive young stellar objects
(YSOs) were studied with single-dish telescopes, to investigate their
physical properties through the analysis of their (sub)mm continuum
and molecular emission
\citep[e.g.,][]{1996A&A...308..573M,1998A&A...336..339M,2000A&A...355..617M,1997MNRAS.291..261W,1998MNRAS.301..640W,1999MNRAS.309..905W,2000A&A...357..637H,2001ApJ...552L.167Z,2005ApJ...625..864Z,2002ApJ...566..931S,2002ApJ...566..945B,2002A&A...383..892B,2004A&A...426...97F,2004A&A...417..115W,2005A&A...434..257W}. However,
an intrinsic feature of high-mass stars is that they form in clusters,
and that most of them are at large (several kpc)
distances. Therefore, single-dish studies, as valuable as they are, lack
the necessary spatial resolution to resolve single protostars and
study the inner regions where high-mass star formation takes place.
Interferometric observations started shedding light into the complex
nature of high-mass star forming regions with the adequate spatial
resolution
\citep[e.g.,][]{1997A&A...325..725C,1999A&A...345..949C,2005A&A...434.1039C,1999ApJ...514L..43W,2002A&A...387..931B,2005ApJ...628..800B,2006ApJ...649..888H,2000ApJ...535..833S}.
However, the number of massive YSOs studied at high resolution is
still too small to establish the general properties of the dense
cores where massive stars form on a statistical base.

In this paper, we present an interferometric analysis of the high-mass
star forming region \object{IRAS\,05358+3543} at (sub)mm wavelengths
in several molecular transitions.  IRAS\,05358+3543 (also known in
literature as \object{S233IR}) is part of a sample of 69 high-mass
protostellar objects studied in great detail in recent years
\citep{2002ApJ...566..931S,2002ApJ...566..945B,2002A&A...383..892B,2002A&A...390..289B,2004A&A...417..115W,2005A&A...434..257W,2005A&A...442..949F}.
At a distance of 1.8~kpc \citep{1990ApJ...352..139S}, IRAS 05358+3543
has a bolometric luminosity of 6300~L$_{\sun}$; strong high-mass star
formation activity is evidenced by maser emission
\citep[see][]{1991ApJ...380L..75M,1995A&AS...112..299T,2000A&A...362.1093M}
and outflow activity \citep{1990ApJ...352..139S,2002A&A...387..931B}.
Previous interferometric observations by \citet{2002A&A...387..931B}
resolved three dust condensations (mm1, mm2 and mm3) within an area of
$9'' \times 4''$ ($\sim17\,100\times7\,200$~AU), and revealed at least
three outflows in CO and SiO, the most prominent of which is more than
a parsec in length, and massive $(M >10~M_ {\sun})$.  Two of the three
identified outflows originate from the vicinity of mm1, which is
probably the main powerhouse in the region.

To zoom in on the innermost region around the protostars, and study the
physical properties of the individual potentially star-forming cores, we carried out a
comprehensive program to observe the region at high spatial
resolution with the Plateau de Bure Interferometer\footnote{IRAM is
supported by INSU/CNRS (France), MPG (Germany) and IGN (Spain).} at 97~GHz and
241~GHz, and
the Submillimeter Array\footnote{The Submillimeter Array is a joint
project between the Smithsonian Astrophysical Observatory and the
Academia Sinica Institute of Astronomy and Astrophysics and is funded
by the Smithsonian Institution and the Academia Sinica.} at 338~GHz.  The new observations reach
a resolution down to ~$0.6''$, corresponding to $\sim 1100$~AU at the
distance of the source. 
\citet{05358-beuther} studied the continuum emission of this dataset. They identified four compact protostellar sources in
the region; mm1 is resolved into two continuum peaks, mm1a and mm1b,
with a projected linear separation of $\sim 1700$~AU.  A mid-infrared
source \citep{2006MNRAS.369.1196L}, and a compact 3.6~cm continuum
source \citep{05358-beuther} coincide with mm1a, which is also
associated with the class II methanol masers detected by
\citet{2000A&A...362.1093M}. The previously identified source mm2
resolves into several sub-sources; however, only one of them (mm2a,
according to the nomenclature used by \citealt{05358-beuther}) is a
protostellar source, while the others are probably caused by the
outflows in the region. The third source mm3 remains a single compact
core even at the highest spatial resolution.
In Table~\ref{cores}, we report the positions of the four
sources identified in the continuum emission by \citet{05358-beuther}.

In this paper, we discuss the spectral line observations complementing
the continuum data discussed by \citet{05358-beuther}. In section \S\ref{observations}, the different observations
are presented.  In section
\S\ref{obs-res}, we discuss our results, and
analyse the extended emission of low excitation molecular transitions
(\S\ref{extended}), as well as the molecular spectra at the positions
of the dust condensations(\S\ref{proto}). Finally, in section
\S\ref{analysis} we derive the physical parameters of the gas around
the protostars from the analysis of their spectra.
In the following sections, we use the term protostar for young
massive stellar objects which are still accreting material from the
surroundings, independently whether they already started burning
hydrogen or not.
\begin{table}
\centering
\caption{Positions of the four dust condensations in IRAS\,05358+3543 \citep[from][]{05358-beuther}.\label{cores}}
\begin{tabular}{lcc}
\multicolumn{1}{c}{Source} &\multicolumn{1}{c}{R.A. [J2000]}&\multicolumn{1}{c}{Dec. [J2000]}\\
\hline
\hline
mm1a&05:39:13.08&35:45:51.3\\
mm1b&05:39:13.13&35:45:50.8\\
mm2a&05:39:12.76&35:45:51.3\\
mm3&05:39:12.50&35:45:54.9\\

\hline
\hline
\end{tabular}
\end{table}

\section{Observations}\label{observations}

\begin{table*}
\centering
\caption{Observational parameters.\label{obs}}
\begin{tabular}{lrccrcr}
\multicolumn{1}{c}{Main line} &\multicolumn{1}{c}{Centre frequency}&\multicolumn{1}{c}{Configuration}&\multicolumn{1}{c}{HPBW}&\multicolumn{1}{c}{P.A.}&\multicolumn{1}{c}{$\Delta v$}&\multicolumn{1}{c}{r.m.s.}\\
& \multicolumn{1}{c}{(GHz)}&&
\multicolumn{1}{c}{($\arcsec$)}&\multicolumn{1}{c}{($^\circ$)}&
\multicolumn{1}{c}{(${\rm km~s}^{-1}$)}&\multicolumn{1}{c}{(Jy/beam)}\\
\hline
\hline
C$^{34}$S $2\to 1$&96.44&BCD&$4.21\times 3.07$&66&0.5&0.01\\
CH$_3$OH $2_{k}\rightarrow 1_{k}~{v}_t=1$&96.51 &BCD&$4.21\times 3.06$&66&0.5&0.01\\
CH$_3$OH $2_{k}\rightarrow 1_{k}~{v}_t=1$&96.51 &AB&$1.85\times 1.36$&26&1.0&0.004\\
CH$_3$OH $2_{k}\rightarrow 1_{k}~v_t=0$&96.72 &BCD&$4.19\times 3.04$&65&8&0.01\\
CH$_3$CN $13_{k}\rightarrow 12_{k}$&238.90    &BCD&$2.12\times 1.26$&77&0.8&0.03\\
C$^{34}$S $5\to 4$ &241.08&AB&$0.77\times 0.55 $&13&1.5&0.01\\
CH$_3$OH $5_{k}\rightarrow 4_{k}~v_t=1$&241.30&AB&$0.77\times 0.55 $&13&1.5&0.01\\
CH$_3$OH $5_{k}\rightarrow 4_{k}~v_t=0$&241.85&BCD&$2.58\times 1.35$&80&0.8&0.03\\
H$_2$CS $7_{1,6}\to 6_{1,5}$ &243.98&AB&$0.77\times 0.55 $&8&1.5&0.01\\
SO$_2~14_{0,14} \to 13_{1,13}$&244.25&AB&$0.77\times 0.55 $&8&1.5&0.01\\
CH$_3$OH $7_{k}\rightarrow 6_{k}~v_t=0,1$&338.72&comp.-ext.&$1.93\times 1.14 $&86&1.0&0.1\\
\hline
\hline
\end{tabular}
\end{table*}

\subsection{Plateau de Bure Interferometer (PdBI)}
\begin{figure*}
\centering
\includegraphics[bb=1 102 802 600,clip,width=15cm]{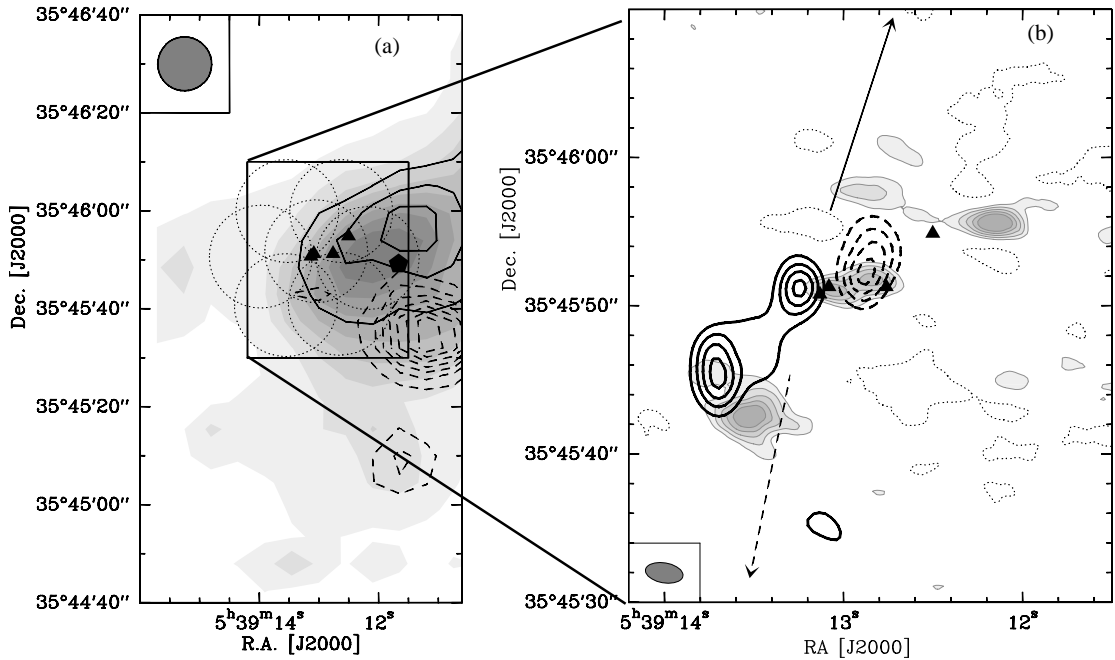}
\caption{{\bf a:)} In grey scale, the map of the integrated intensity
of the $5_{0} \to 4_{0}~v_t=0$~CH$_3$OH-$A$ line with the 30~m IRAM
telescope ($v=[-17,-16]$~km~s$^{-1}$; levels from 0.5~Jy~beam$^{-1}$~km~s$^{-}$ in steps of 0.5). The dashed contours
show the blue-shifted emission ($v=[-23,-20]$~km~s$^{-1}$; levels from
0.5~Jy~beam$^{-1}$~km~s$^{-1}$ in steps of 0.5); the solid
lines the red-shifted emission ($v=[-14,-12]$~km~s$^{-1}$; levels from
1~Jy~beam$^{-1}$~km~s$^{-1}$ in steps of 0.5).  The beam is
indicated in the top left corner. The dotted circles outline the
observed mosaic of seven fields. The pentagon locates the position of
one of the H$^{13}$CO$^+$ peaks observed by
\citet{2002A&A...387..931B}, which is the candidate centre for the
outflow traced by CH$_3$OH. The solid lines outline the region shown
in panel b). {\bf b:)} In grey scale (and grey contours), the map of
the integrated intensity of the $5_{0} \to 4_{0}~v_t=0$~CH$_3$OH-$A$
line with the Plateau de Bure Interferometer
($v=[-24,-12]$~km~s$^{-1}$).  Level contours are from 1~Jy~beam$^{-1}$~km~s$^{-1}$, in step of 0.5.  Dotted contours show the negative emission (-0.8~Jy~beam$^{-1}$~km~s$^{-1}$). 
The black contours outline the high
velocity outflow (CO $2\to1$ SMA data, Beuther
priv. comm. Blue-shifted emission $v=[-44,-24]$~km~s$^{-1}$;
red-shifted emission $v=[-8,2]$~km~s$^{-1}$; levels from 30 to 100
~Jy~beam$^{-1}$~km~s$^{-1}$ in steps of 20). The four triangles mark the positions of the
main mm dust condensations.  For illustration, the direction of the highly collimated outflow
 \citep{2002A&A...387..931B} is shown by the arrows. For both flows, dashed lines are for the
blue-shifted emission, solid lines for the red-shifted
emission.
 The beam is indicated in
the bottom left corner.}\label{790}

\end{figure*}

IRAS 05358+3543 was observed with the IRAM Plateau de Bure
Interferometer in two different frequency setups, in 2003 and 2005.  A
first frequency setup was performed in four tracks between January and
October 2003 in the  BC and D configurations of the array.  Two
observations, in January and October, were performed with only 5
antennas; on January 12th, 2003, the array was configured in a special
BC combination.  The 3~mm receivers were used in
single-side band mode and tuned to 96.6~GHz; the 1~mm receivers, in double-side band mode, were tuned to 241.85~GHz (USB). At 3~mm, the C$^{34}$S line and the
torsionally excited $2_k \rightarrow 1_k$ CH$_3$OH quartet were covered using
two correlator units of 80~MHz bandwidth. One 320~MHz unit was placed
to obtain a continuum measurement at 3~mm. The  $5_k\rightarrow
4_k~\rm{v_t=0}$ CH$_3$OH band, the  $2_{1,1}\rightarrow 2_{1,2}$ HDO and the
$5\rightarrow 4$ SO$_2$ lines were observed with four units of 160~MHz
bandwidth, which were also used to obtain a continuum measurement at
1.3mm. The same configuration of the correlator units allowed the
observation of $13_k\rightarrow 12_k$  CH$_3$CN band in the
LSB at 239~GHz.
The observations were performed in the MOSAIC mode, with seven fields
covering the whole source in a hexagonal pattern with a centre field (see Fig.~\ref{790}).

 The second
frequency setup was observed on February 4th and February 7th, 2005,
in the A and B configurations.  The 3~mm receivers were used in single-side 
band mode and tuned again to 96.6~GHz, with the same configuration of the
correlator 
described before. 
The 1~mm receivers, in double
side-band mode, were tuned to 241.2~GHz (USB) with three 160~MHz units covering
the $5_{k}\rightarrow 4_{k}~\rm{v_t=1}$ CH$_3$OH band.

Bandpass calibration was done with 0420-014, 3C454.2 and NRAO150
for the BCD configuration, with 3C84 for the AB data. NRAO150,
0528+134 and 3C273 were used as flux calibrators of the BCD and AB
data, respectively.  Phase and amplitude calibration was done via
observations of 0528+134, 0552+398, 0529+483 and J0418+380.  Measured
system temperatures were between 100 and 260~K in the 3.1~mm band for
both setups.  On February 4th, the system temperatures in the 1.2~mm
receivers ranged between 240 and 420~K, with one receiver measuring
system temperatures of 900~K. On February 7th, and during the
observations of the first frequency setup, the system temperatures in
the 1.2~mm bands were higher, between 400 and 1000~K, due to less
favourable weather conditions.

For both frequency setups, the phase centre was
$\alpha_{2000}$=05$^h$39$^m$13$^s$.07,
$\delta_{2000}$=$+$35$^{\circ}$45$'$50$''$.5, with
v$_{\rm{LSR}}$=-17.6 km s$^{-1}$.  Details on the spectral
resolutions, the synthesised beam sizes and the main spectral
lines per frequency units are given in Table~\ref{obs}. For the AB configuration the
baselines range between 30 and 400~m, for the BCD configuration
between 20 and 320~m. Therefore, at 1.2~mm any source structure larger
than 6$''$ is filtered out in the AB observations, and larger
than 13$''$ in the BCD data.

The data calibration and the imaging were performed with the CLIC and
MAPPING software\footnote{http://www.iram.fr/IRAMFR/GILDAS}. For
the PdBI and the SMA data, line free channels were averaged to produce
continuum images, which were then subtracted from the line data in the
visibility plane.\footnote{The calibrated UV tables related to the PdBI and SMA observations are 
available in electronic form at the CDS. The integrated intensity map of the $5_{0} \to 4_{0}~v_t=0$~CH$_3$OH-$A$ line with the 30~m IRAM is also available in electronic form at the CDS.}

\subsection{Submillimeter Array}
We observed IRAS\,05358+3543 with the SMA on November 11th, 2004, at
348\,GHz (865\,$\mu$m) in the compact configuration with seven
antennas, and on January 15th, 2005, in the extended configuration
again with seven antennas in the array. However, the data from one
antenna in the extended configuration were unusable reducing it to six
effective antennas for that configuration. Due to problems at the correlator 
during the observations in the extended
configuration, the frequency of the $7\to6$ C$^{34}$S line was not covered. 
However, the $7\to6$ C$^{34}$S map was produced from the compact configuration data.
 The projected baselines
ranged between   13 and 223~m. The short baseline cutoff
implies that source structures $\geq 16''$ are filtered out by the
observations. The phase centre of the observations was
$\alpha_{2000}$=05$^h$39$^m$13$^s$.07 and
$\delta_{2000}$=$+$35$^{\circ}$45$'$51$''$.2 with a
$v_{\rm{lsr}}=-17.6$\,km\,s$^{-1}$. Bandpass calibration was done with
Jupiter, Uranus, Callisto, and 3C279. We used Callisto and 3C279 for
the flux calibration which is estimated to be accurate within
20\%. Phase and amplitude calibration was done via frequent
observations of the quasar 3C111 about $16.3^{\circ}$ from the phase
centre. The zenith opacities, measured with the NRAO tipping
radiometer located at the Caltech Submillimeter Observatory, were good
during both tracks with $\tau(\rm{348GHz})\sim 0.18$ (scaled from the
225\,GHz measurement via $\tau(\rm{348GHz})\sim 2.8\times
\tau(\rm{225GHz})$). The receiver operated in a double-side band mode
with an intermediate frequency of 4--6\,GHz so that the upper and lower
side band were separated by 10\,GHz. The correlator had a bandwidth of
2\,GHz and the channel separation was 0.8125\,MHz. Measured
double-side band system temperatures corrected to the top of the
atmosphere were between 150 and 500\,K, mainly depending on the
elevation of the source.  Details on the observational setup are given in Table~\ref{obs}.

The initial flagging and calibration was done with the IDL superset
MIR originally developed for the Owens Valley Radio Observatory
and adapted for the SMA\footnote{The MIR
cookbook by Charlie Qi can be found at
http://cfa-www.harvard.edu/$\sim$cqi/mircook.html.}. The imaging and
data analysis was conducted in MIRIAD
\citep{1995ASPC...77..433S} and MAPPING. During the observations on November 11th,
2004, the position of the primary calibrator 3C111 was wrong in the
catalogue by $\sim$0.6$''$.  Therefore, we self-calibrated our
secondary phase calibrator 0552+398 ($16.3^{\circ}$ from the source)
shifting it in the map to the correct position. The solutions were
then applied to IRAS\,05358+3543.

\subsection{Single-dish observations with the IRAM 30~m telescope}
In addition to the high resolution data, we mapped an area of
$70''\times180''$ in the $5_k \to 4_k~v_{\rm t}=0$ CH$_3$OH band, with
the HERA receiver \citep{2004A&A...423.1171S} at the IRAM 30~m telescope in on-the-fly mode. The
observations were performed in service-mode in February 2005, under
excellent weather conditions (0.6--1.1~mm precipitable water
vapour). The pointing was checked on Saturn and on a nearby source
(0439+360) and was found to be accurate to $\sim 6''$. Conversion from
antenna temperature to main-beam brightness temperature was performed by using a
beam efficiency of 0.48\footnote{http://www.iram.fr/IRAMES/}.  The
beam of the 30~m telescope at 241.8~GHz is
$\sim10.2''$. Unfortunately,  the overlapping in the UV
plane of our PdBI and 30~m data is poor, since the PdBI baselines  start only only at 20~m, 
and we were not able to combine the two
datasets and recover the short spacing information. The IRAM 30~m data are, however, used
in the following discussion to study the extended structure of the methanol emission.

\section{Observational results}\label{obs-res}
\renewcommand{\baselinestretch}{1.1}
\begin{table}
\centering
\caption{Table of CH$_3$OH detected transitions. In Col.~4,  $\star$ is used to indicate the lines with extended emission; (?)
for tentative detections.}\label{ch3oh-lines}
\begin{tabular}{lrrc}
\multicolumn{1}{c}{Transition} &\multicolumn{1}{c}{Rest frequency}&\multicolumn{1}{c}{$E_{\rm{upper}}$}&\multicolumn{1}{c}{Detections$^a$}\\
&\multicolumn{1}{c}{(GHz)}&\multicolumn{1}{c}{(K)}\\
\hline
\hline

$E$ $2_{1} \to 1_{1}~v_t=1$&96.492&298&1a\\
$E$ $2_{0} \to 1_{0}~v_t=1$&96.493&308&1a\\
$A$ $2_{0} \to 1_{0}~v_t=1$&96.514&431&1a(?)\\

$E$ $2_{-1} \to 1_{-1}~v_t=0$&96.739$^b$&13&1,2,3\\
$A$ $2_{0} \to 1_{0}~v_t=0$&96.741$^b$&7&1,2,3\\
$E$ $2_{0} \to 1_{0}~v_t=0$&96.745$^b$&20&1,2,3\\
$E$ $2_{1} \to 1_{1}~v_t=0$&96.756$^b$&28&1,2,3\\

$E$ $22_{-6}\rightarrow 23_{-5}~v_t=0$&241.043&776&1a(?)\\

$E$ $5_{3} \to  4_{3}~v_t=1 $& 241.167&452&1a\\
$A$ $5_{\pm 4}  \to 4_{\pm  4}~v_t=1 $& 241.178&516&1a\\
$E$ $5_{-3}\to 4_{-3}~v_t=1$& 241.180&357&1a\\
$E$ $5_{-4}\to 4_{-4}~v_t=1$& 241.184&440&1a\\
$E$ $5_{-2}\to 4_{-2}~v_t=1$& 241.187&399&1a\\
$A$ $5_{2} \to  4_{2}~v_t=1 $& 241.193&333&1a\\
$A$ $5_{2} \to  4_{2}~v_t=1 $& 241.196&333&1a\\
$A$ $5_{\pm 3} \to  4_{\pm 3}~v_t=1 $& 241.198&431&1a\\
$E$ $5_{1} \to  4_{1}~v_t=1 $& 241.204&326&1a\\
$E$ $5_{0} \to  4_{0}~v_t=1 $& 241.206&335&1a\\
$E$ $5_{2} \to  4_{2}~v_t=1 $& 241.211&435&1a\\

$E$ $5_{-1} \to 4_{-1}~v_t=1$&241.238&448&1a\\
$A$ $5_{0} \to 4_{0}~v_t=1$&241.268&458&1a\\

$A$ $5_{1} \to 4_{1}~v_t=1$&241.441&360&1a\\

$E$ $5_{0}   \to 4_{0}~v_t=0   $& 241.700&48&1,2,3,line,$^\star$ \\
$E$ $5_{-1} \to 4_{-1 }~v_t=0 $& 241.767&40&1,2,3,line,$^\star$ \\
$A$ $5_{0}   \to 4_{0}~v_t=0   $& 241.791&34&1,2,3,line ,$^\star$\\
$A$ $5_{\pm 4}   \to 4_{\pm 4}~v_t=0   $& 241.807&115&1,2,line \\
$E$ $5_{-4}  \to 4_{-4}~v_t=0  $& 241.813&123&1,2,line \\
$E$ $5_{4}   \to 4_{4}~v_t=0   $& 241.830&131&1,2,line \\
$A$ $5_{\pm 3}   \to 4_{\pm 3}~v_t=0   $& 241.833&85&1,2,line \\
$A$ $5_{2}   \to 4_{2}~v_t=0   $& 241.842$^c$&73&1,2,line \\
$E$ $5_{3}   \to 4_{3}~v_t=0   $& 241.844$^c$&83&1,2,line \\
$E$ $5_{-3}  \to 4_{-3}~v_t=0  $& 241.852&98&1,2,line \\
$E$ $5_{1}   \to 4_{1}~v_t=0   $& 241.879&56&1,2, 3(?),line \\
$A$ $5_{2}   \to 4_{2}~v_t=0   $& 241.888&73&1,2,line \\
$E$ $5_{-2}  \to 4_{-2}~v_t=0  $& 241.904$^c$&61&1,2,3,line,$^\star$ \\
$E$ $5_{2}   \to 4_{2}~v_t=0   $& 241.905$^c$&57&1,2,3,line,$^\star$ \\

$A$ $5_{1} \to  4_{1}~v_t=0 $&243.916&40&1,2,line,$^\star$\\

$E$ $7_{-2}   \to 6_{-2,v_=1}   $&337.605&429&1a  \\
$A$ $7_{2}    \to 6_{2}~v_t=1    $&337.636&363&1a \\
$E$ $7_{-1}   \to 6_{-1}~v_t=1   $&337.642&356&1a \\
$E$ $7_{0}    \to 6_{0}~v_t=1    $&337.644&365&1a  \\
$E$ $7_{-4}   \to 6_{-4}~v_t=1   $&337.646&470&1a \\

$E$ $7_{0}    \to 6_{0}~v_t=0    $&338.125&78&1,2,line,$^\star$  \\
$E$ $7_{-1}   \to 6_{-1}~v_t=0   $&338.345&71&1,2, 3(?),line,$^\star$  \\
$E$ $7_{6}    \to 6_{6}~v_t=0   $&338.405&244&1  \\
$A$ $7_{0}    \to 6_{0}~v_t=0   $&338.409&65&1,2, 3(?),line,$^\star$ \\
$A$ $7_{\pm 5}    \to 6_{\pm 5}~v_t=0   $&338.486&203 &1 \\
$A$ $7_{\pm 4}    \to 6_{\pm 4}~v_t=0   $&338.513$^c$&145&1,2,line  \\
$A$ $7_{2}    \to 6_{2}~v_t=0   $&338.513$^c$&103&1,2,line  \\
$E$ $7_{4}    \to 6_{4}~v_t=0   $&338.530&161&1  \\
$A$ $7_{3}    \to 6_{3}~v_t=0   $&338.541$^c$&115&1,2,line  \\
$A$ $7_{3}    \to 6_{3}~v_t=0   $&338.543$^c$&115&1,2,line   \\
$E$ $7_{-3}   \to 6_{-3}~v_t=0  $&338.560&128&1  \\
$E$ $7_{3}    \to 6_{3}~v_t=0   $&338.583&118&1,2,line  \\
$E$ $7_{1}    \to 6_{1}~v_t=0   $&338.615$^d$&86&1,2,line  \\
$A$ $7_{2}    \to 6_{2}~v_t=0   $&338.640 &103&1,2,line \\
$E$ $7_{2}    \to 6_{2}~v_t=0   $&338.722$^c$&87&1,2,line,$^\star$ \\
$E$ $7_{-2}   \to 6_{-2}~v_t=0   $&338.723$^c$&91&1,2,line,$^\star$ \\
\hline
\hline
\end{tabular}
\begin{list}{}{}
\item $^a$ 1,2,3 indicate mm1, mm2, and mm3, respectively. 1 is used for data where mm1a and mm1b are not resolved; line is used
for the position (1.2$''$,0.6$''$) from mm2.
\item $^b$ the $2_k \to 1_k$ quartet of lines is unresolved in velocity;
\item $^c$ blend of lines;
\item $^d$blend with the $20_{1,19}\to 19_{2,18}$~SO$_2$ line at       338.612~GHz.

\end{list}
\end{table}
\renewcommand{\baselinestretch}{1.5}
In this section we present the results of the line observations
 performed towards IRAS\,05358+3543. Since the observations were aimed at observing
several bands of methanol transitions, the majority of detected lines comes from this molecule. However,
emission from other molecular species with transitions close in  frequency space to methanol is also detected.
All detected lines
are listed in Tables~\ref{ch3oh-lines} (for CH$_3$OH) and \ref{lines} (for the other molecular species); the dust condensations where the lines are
detected are also reported in the tables.

\begin{table}
\centering
\caption{Table of observed transitions from other molecular species. n Col.~4,  $\star$ is used to indicate the lines with extended emission. \label{lines}}
\begin{tabular}{lrrc}

\multicolumn{1}{c}{Transition} &\multicolumn{1}{c}{Rest frequency}&\multicolumn{1}{c}{$E_{\rm{upper}}$}&\multicolumn{1}{c}{Detections$^a$}\\
&\multicolumn{1}{c}{(GHz)}&\multicolumn{1}{c}{(K)}\\
\hline
\hline
C$^{34}$S $2\to 1$&96.413&7&1,2,3,$\star$\\
C$^{34}$S $5\to 4$&241.016&35&1\\
CH$_3$CN $13_{6} \to 12_{6}$& 238.972&337&1\\
CH$_3$CN $13_{5} \to 12_{5}$& 239.022&259&1\\
CH$_3$CN $13_{4} \to 12_{4}$& 239.065&195&1\\
CH$_3$CN $13_{3} \to 12_{3}$& 239.096&145&1,line\\
CH$_3$CN $13_{2} \to 12_{2}$& 239.119&109&1\\
CH$_3$CN $13_{1} \to 12_{1}$& 239.133&87&1,line\\
CH$_3$CN $13_{0} \to 12_{0}$& 239.138&80&1,line\\

HDO $2_{1,1}\rightarrow 2_{1,2}$&241.561&95&1\\
SO$_2$ $5_{2,4}\rightarrow 4_{1,3}$&241.616&24&1,line\\ 
HNCO $11_{2,10} \to 10_{2,9}$&241.704&243&1\\
HNCO $11_{2,9} \to 10_{2,8}$&241.708&243&1\\
HNCO $11_{0,11} \to 10_{0,10}$&241.774&70&1\\

H$_2$CS  $7_{1,6}\to 6_{1,5}$& 244.048&60&1\\
SO$_2$ $14_{0,14}\rightarrow 13_{1,13}$&244.254&94&1\\
C$^{34}$S $7\to 6$&337.396&65&1,line\\
H$_2$CS  $10_{1,10} \to 9_{1,9}$& 338.083&102&1\\
SO$_2$ $18_{4,14}\rightarrow 18_{3,15}$&338.306&197&1\\
SO$_2$ $20_{1,19}\rightarrow 19_{2,18}$&338.612$^b$&199&1\\
\hline
\hline
\end{tabular}
\begin{list}{}{}
\item $^a$ 1,2,3 indicate mm1, mm2, and mm3, respectively. 1 is used
for data where mm1a and mm1b are not resolved; line is used for the
position (1.2$''$,0.6$''$) off of mm2.
\item $^b$ blend with the $7_{1}\to 6_{1}~v_t=0$~CH$_3$OH-$E$ line at  338.615~GHz
\end{list}
\end{table}

The large structured emission is sampled by the single dish methanol 
data, while the  interferometric datasets allow us to zoom in on the gas around the protostars. 
These data range in angular
 resolution from $4''$ to $0.6''$, corresponding to 7200 and 1100~AU,
 respectively. Another property of these data is  that a broad range of structures  are
 filtered out from the observations, due to the missing short-spacing
 flux, from $\sim25''$ in the BCD configuration of the PdBI at 3~mm to
 $\sim6''$ in its AB configuration at 1.3~mm. This means that the
 various datasets are differently affected by the problem, and the
 comparison of transitions from different configurations is not
 straightforward.  Moreover, the UV coverage is poorly
 sampled also at intermediate scales, and smaller structures can be
 filtered out. Comparing the single dish data for the $5_k\to 4_k~v_t=0$ band of CH$_3$OH 
to the interferometric observation confirms that the PdBI data are missing fluxes.
Although we cannot perform the same comparison for the rest of the dataset, 
negative features due to  missing short spacings, heavily corrupt the other
low  energy
 line datacubes.
 Thus, the interpretation of the line interferometric data is 
 potentially affected by missing flux problems.

Although our observations are among the highest spatial
 resolution studies of high-mass star forming regions existing today,
 there are obviously still limitations to what they resolve. We know
 from other observations that multiplicity on smaller scales happens
 in star forming regions of all masses
 \citep[e.g.,][]{1999A&A...347L..15W,2005ApJ...622L.141M}. 
 The highest linear resolution of our observations  ($\sim 1100$~AU) is reached only in
 one dataset, while the typical resolution of the other data is
 poorer. Within our own data, the two sources with the smallest
 separation, mm1b and mm1a, are spatially resolved only in one
 dataset. Therefore, we cannot exclude that higher resolution 
would reveal more objects, and that
 the parameters we derive do not sample the gas around a single protostar.
 However, if this was the case, the multiple unresolved
 sources would form from the same reservoir of dust and gas. Assuming
 that they are gravitationally bound, the parameters derived in our current study would
 still be relevant to the analysis of the collapsing cores.

\subsection{Extended emission}\label{extended}

Emission from relatively low excitation lines is extended, as shown in
Fig.~1(a), where the integrated intensity of the $5_{0} \to 4_{0},
v_t=0$~CH$_3$OH-$A$ transition taken with the 30~m telescope is
presented.

For methanol, the emission at the cloud velocity shows a distribution
very similar to the continuum emission.  The offset of the line emission peak from the continuum peak is
real, since it was already observed  in
methanol and in other molecular species \citep{2002A&A...387..931B}. In their single-dish
observations, \citet{2002A&A...387..931B} detected red and blue non-Gaussian emission in the $5_k \to 4_k$ methanol band, and suggested
that both emissions are associated with a third outflow in the region to the
west of the continuum peak.  Their results are confirmed by our data,
which show a bipolar
distribution in the low excitation transitions of CH$_3$OH (Fig.~1(a)) with a better signal-to-noise ratio.
Within  
the pointing uncertainties and the low resolution of the data, the centre of this outflow seems to be associated
with one of the H$^{13}$CO$^{+}$ peaks (indicated by a pentagon in
Fig.~1(a)) detected by \citet{2002A&A...387..931B}. If
this were the case, our observations would solve the problem of the
powering source of the third outflow of the region ($C$ in the notation of \citealt{2002A&A...387..931B}), 
which was previously not
assigned.  However, higher angular resolution is needed to 
assign the centre of this flow beyond a doubt.

The interpretation of the interferometric data of the same transition
is not as straightforward as for the single-dish data.  In Fig.~1(b), the integrated intensity of the $5_{0}
\to 4_{0}$~CH$_3$OH-$A$ line is shown as seen in the PdBI observations,
together with the direction of the two outflows originating from the
vicinity of mm1. The region west
of the dust cores, where the bipolar distribution is detected in the
single-dish data, is only partially in the field of view of the PdBI
observations.  
The intensity map is strongly affected by
missing flux problems, which results in filtering out the extended
emission. For this reason, the PdBI map of this transition looks
clumpy.  The emission is detected all over the dust continuum peaks,
often red- and/or blue-shifted.
These
emission spots could be caused by the high velocity CO outflow
originating from the vicinity of mm1 (see Fig.1(b)).  However, we refrain from a further discussion of these
features, given the missing short
spacing information of our methanol data.

\begin{figure}
\centering
\includegraphics[width=8cm]{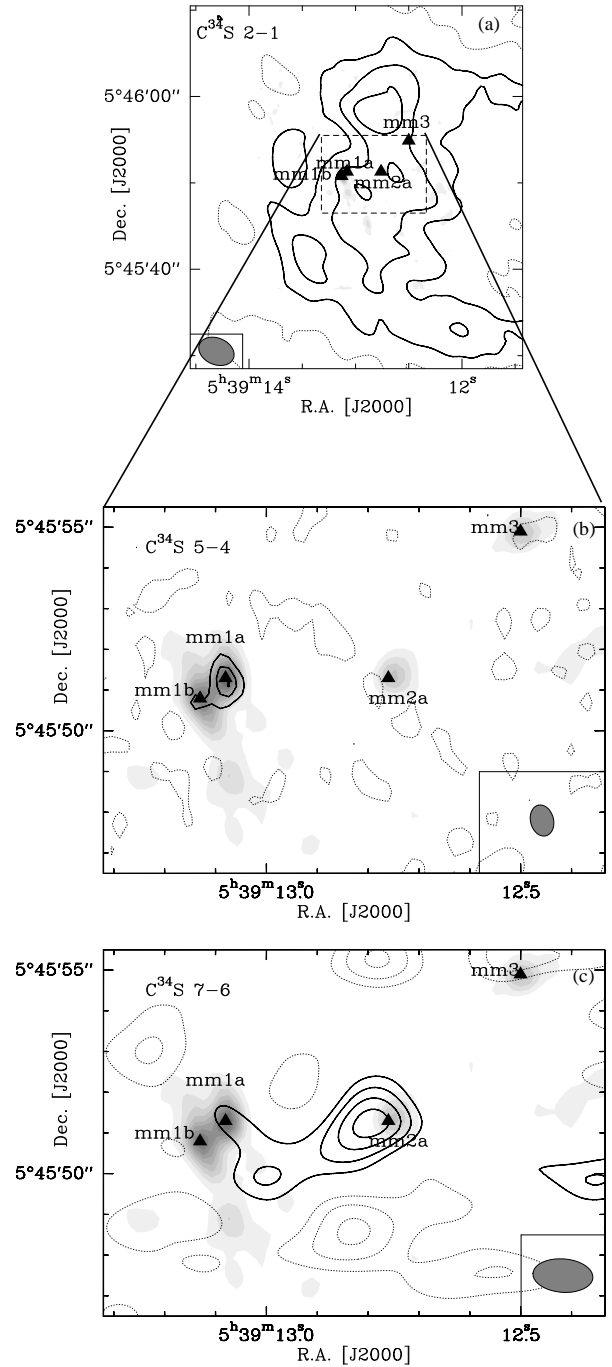}
\caption{In grey scale the 1.3~mm continuum emission (from the AB
configuration of the Plateau de Bure Interferometer). The black
contours show the integrated line intensity of the $2\to 1$
({\bf a}), the $5\to 4$ ({\bf b}), and of the $7\to 6$~C$^{34}$S
lines ({\bf c}).  Level contours are from 0.1~Jy~beam$^{-1}$~km~s$^{-1}$ in steps of 0.2 for the $2\to 1$ transition; from
0.3~Jy~beam$^{-1}$~km~s$^{-1}$ in steps of 0.3 for the $5\to 4$
line, and from 1.5 in steps of 0.5 for the $7\to 6$. Level contours for the continuum are
in step of 0.003 from 0.003~Jy~beam$^{-1}$. The dotted contours show the negative emission (-0.1~Jy~beam$^{-1}$~km~s$^{-1}$ for the $2\to 1$
 and $5\to 4$ lines; from -1.5 in step of 0.5~Jy~beam$^{-1}$~km~s$^{-1}$ for the $7\to 6$ transition). The dashed
lines in panel a) outline the region shown in the other two panels.   The
different beams are also given.}\label{cs}
\end{figure}

The distribution of C$^{34}$S line is puzzling (Fig.~\ref{cs}). While the $2\to 1$ line
extends over an area of approximately $20''\times 20''$ around the main
dust condensations, and the $7\to 6$ transition is found in
association with mm1 (although marginally detected) and towards mm2, the $5\to 4$ line is detected
 only on mm1a and mm1b (see Fig.~\ref{cs}). In the $5\to
4$~C$^{34}$S line, two velocity components (see discussion on mm1 in \S\ref{proto}) are
detected, one at v$_{lsr}\sim-17.6$~km~s$^{-1}$ and the other at
v$_{lsr}\sim-13$~km~s$^{-1}$, the first on mm1a and the second on
mm1a and on the vicinity of mm1b.   The emission of the $5\to 4$~C$^{34}$S line
extends to the south-west, as does that of the $7\to 6$ line. However, the
spectra are noisy, and the line is detected with a significance level
not better than 2~$\sigma$. In Fig.~\ref{cs}, the level contours start from a value of 3~$\sigma$, and the feature on the south-west of mm1a is not seen in the 
$5\to 4$ transition. 

\citet{1999A&A...352..266O} studied the
excitation of CS and C$^{34}$S towards a sample of ultra-compact H{\sc
ii} regions, and found, as expected given the energies of the levels
involved, that the $5\to 4$~C$^{34}$S transition is usually stronger
than the $7\to 6$ line, making the non detection of the $5\to 4$ line on mm2 difficult to interpret.  
However, this line was observed in the
AB configurations of the PdB interferometer, where source structures
 larger than $ 6''$ are filtered out. Therefore, the unexpected spatial
distribution of this transition, as well as its profile, can be an artifact due to the missing
short spacing information. 

\begin{figure*}
\centering
\includegraphics[angle=-90,width=15cm]{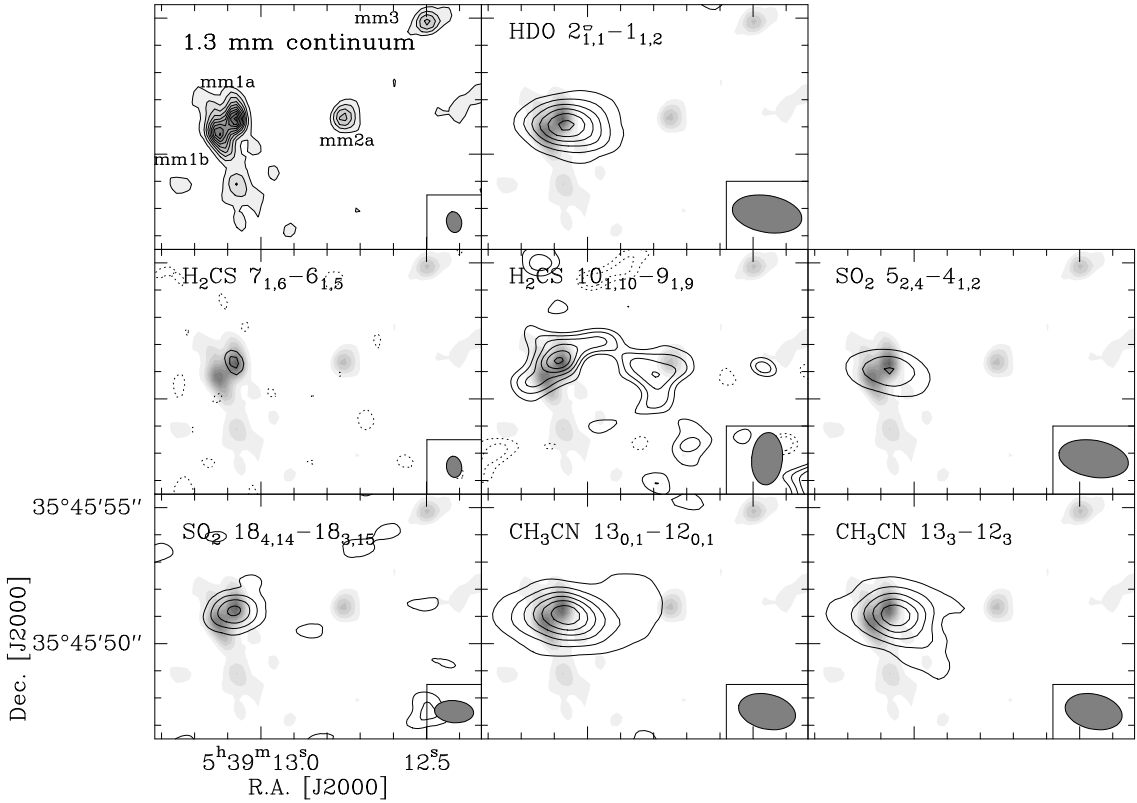}
\caption{Compilation of integrated intensity maps of transitions from
several molecular species, shown as black contours.  In grey scale
(and in the solid contours in the first panel) the 1.3~mm continuum
emission from the AB configuration of the Plateau de Bure
Interferometer. Level contours for the line images are from
0.5~Jy~beam$^{-1}$~km~s$^{-1}$, in step of 0.3; for the $7_{1,6}\to
6_{1,5}$~H$_2$CS transition, from
0.3~Jy~beam$^{-1}$~km~s$^{-1}$, and from 1~Jy~beam$^{-1}$~km~s$^{-1}$ for the 18$_{4,14}\to 18_{3,15}$~SO$_2$ line. Level contours for the $13_{0,1}\to 12_{0,1}$~CH$_3$CN transitions are in step of 0.6~Jy~beam$^{-1}$~km~s$^{-1}$. The dotted contours show the negative emission: for the $7_{1,6}\to
6_{1,5}$~H$_2$CS transition -0.1~Jy~beam$^{-1}$~km~s$^{-1}$; for the  $10_{1,10}\to
9_{1,9}$~H$_2$CS line -1.5 and -1.2~Jy~beam$^{-1}$~km~s$^{-1}$. Continuum contours are the same as in Fig.~\ref{cs}. The beam is
shown in the right bottom corner. 
}\label{allspecies}
\end{figure*}

Comparing the spatial distribution of the
$2\to 1$ and $7\to 6$ C$^{34}$S lines, which are affected by missing
flux problems on larger scales ($\ge$30$''$ and $\ge$16$''$, respectively),
it emerges that they do not peak on the protostellar cores, but
in the gas between the cores, where the temperature most likely drops
down.  Observations of the hot core in G29.96-0.02
\citep{beuther-g29}, with the same frequency setup of our SMA data, show a similar effect, with the $7\to 6$ C$^{34}$S line peaked in the gas between the (sub)mm 
continuum peaks. On the other hand, in G29.96-0.02, the $8_8\to 7_7$~$^{34}$SO line peaks on the hot core. 
Since CS should desorb from dust grains
relatively early in the evolution of the source
\citep{2004MNRAS.354.1141V}, while the SO and SO$_2$ abundances should
increase with time, \citet{beuther-g29} interpret the distributions of
C$^{34}$S and $^{34}$SO in G29.96-0.02 as an evolutionary indicator. 
In our case, we did not detect the $8_8\to 7_7$~$^{34}$SO transition toward IRAS\,0535+3543. However, we observed
four transitions of SO$_2$, and all peak
on the main dust condensation mm1 (Fig.~\ref{allspecies}). Although the evolutionary scenario
outlined for G29.96-0.02 seems to be confirmed by our observations of the
$2\to 1$ and $7\to 6$ C$^{34}$S, and of the SO$_2$ transitions, the distribution of the  $5\to 4$~C$^{34}$S line
does not find an easy interpretation within this theory.

Similarly to C$^{34}$S, also H$_2$CS shows different morphologies in
the two transitions in our datasets (Fig.~\ref{allspecies}), with the $10_{1,10} \to 9_{1,9}$
line ($E_{up}\sim100$~K) having a second peak at an offset position from mm2 (see the
discussion below), while the $7_{1,6}\to 6_{1,5}$ is found only on
mm1a, despite its lower level energy ($E_{up}\sim60$~K). 
We believe that these data are heavily contaminated by missing flux problems
at intermediate and short scales, and refrain from any analysis of this molecular species.

\subsection{Molecular emission from the dust cores}\label{proto}

\begin{figure}
\resizebox{\hsize}{!}{\includegraphics[angle=-90]{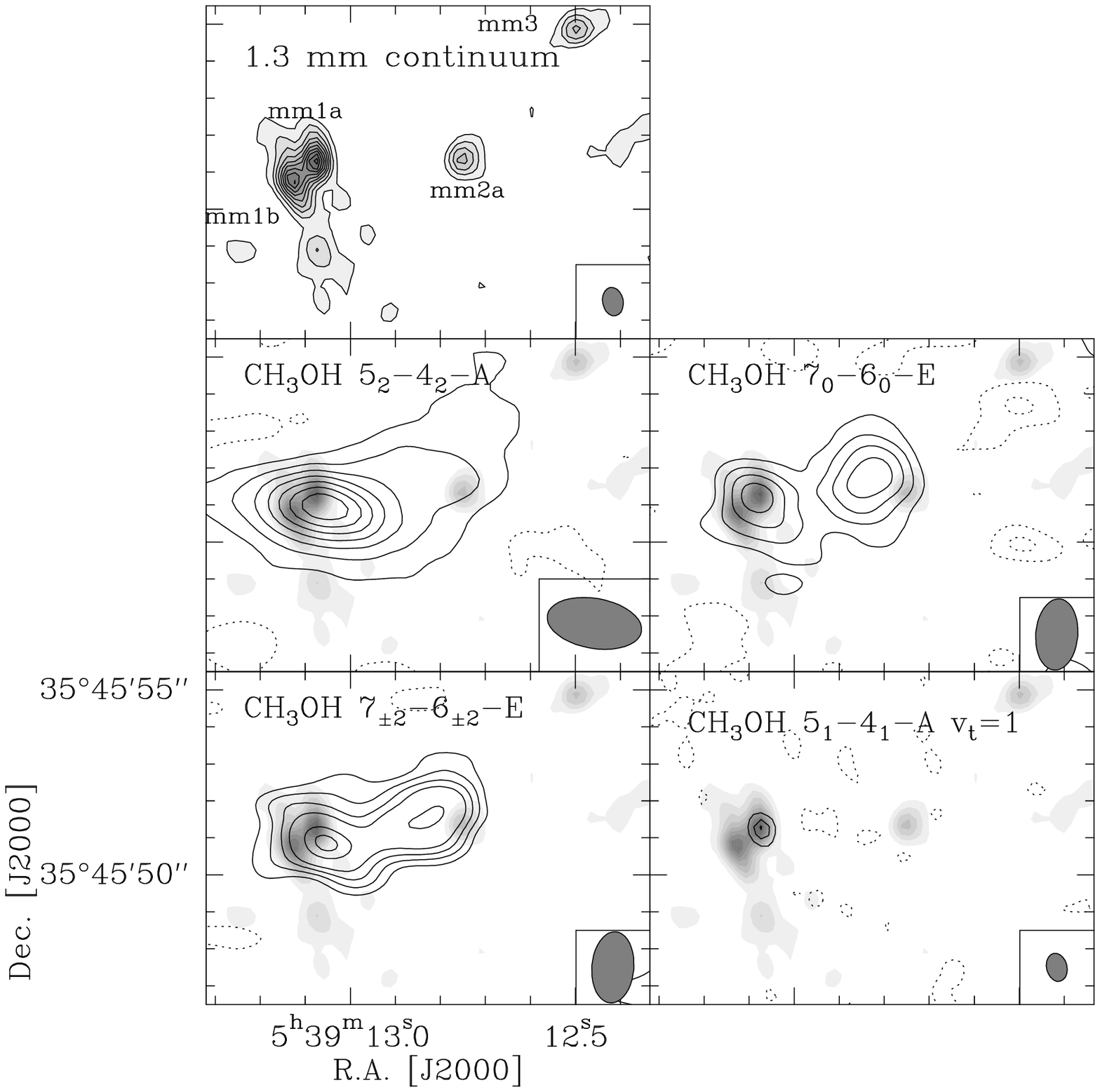}}
\caption{Compilation of integrated intensity maps of CH$_3$OH, shown as black contours.
In grey scale (and in the solid contours in the first panel) the 1.3~mm continuum emission from the AB configuration of the 
Plateau de Bure Interferometer. Level contours for the line images
  are from: 0.3~Jy~beam$^{-1}$~km~s$^{-1}$ in step of 0.4,   1~Jy~beam$^{-1}$~km~s$^{-1}$ in step of 1,
  1.5~Jy~beam$^{-1}$~km~s$^{-1}$ in step of 1,
  0.2~Jy~beam$^{-1}$~km~s$^{-1}$ in step of 0.2, respectively. Dotted contours are used for negative values: for the $5_2 \to 4_2$-$A$ line
  -0.2~Jy~beam$^{-1}$~km~s$^{-1}$; -2 and -1~Jy~beam$^{-1}$~km~s$^{-1}$ for the  $7_0 \to 6_0$-$E$ line; -2~Jy~beam$^{-1}$~km~s$^{-1}$ for the  $7_{\pm 2} \to 6_{\pm 2}$-$E$ line; -0.1 for the $5_1\to4_1$~CH$_3$OH-$E$~$v_t=1$ transition. Continuum contours are the same as in Fig.~\ref{cs}. The beam is shown in the right bottom corner.}\label{ch3oh}
\end{figure}

The molecular spectra of the three main dust condensations, mm1, mm2
and mm3, differ significantly from each other, reflecting the
different physical conditions, and probably time evolution, of the
dust cores.   Maps of the integrated intensity of several transitions
are shown in Figs.~\ref{allspecies} and \ref{ch3oh}. 
All high excitation lines peak on mm1, and specifically
on mm1a for the dataset where the cores mm1a and mm1b are spatially resolved.
However, several moderately excited ($E_{up}\le 130$~K) methanol lines
are extended towards the second dust core mm2, and show a peak of
intensity at (1.2$''$,0.6$''$) from mm2.

\begin{figure*}[]
\centering
\subfigure[]{\label{ch3cn-mm1}
\includegraphics[bb= 200 16 552 681,clip, angle=-90,width=15cm]{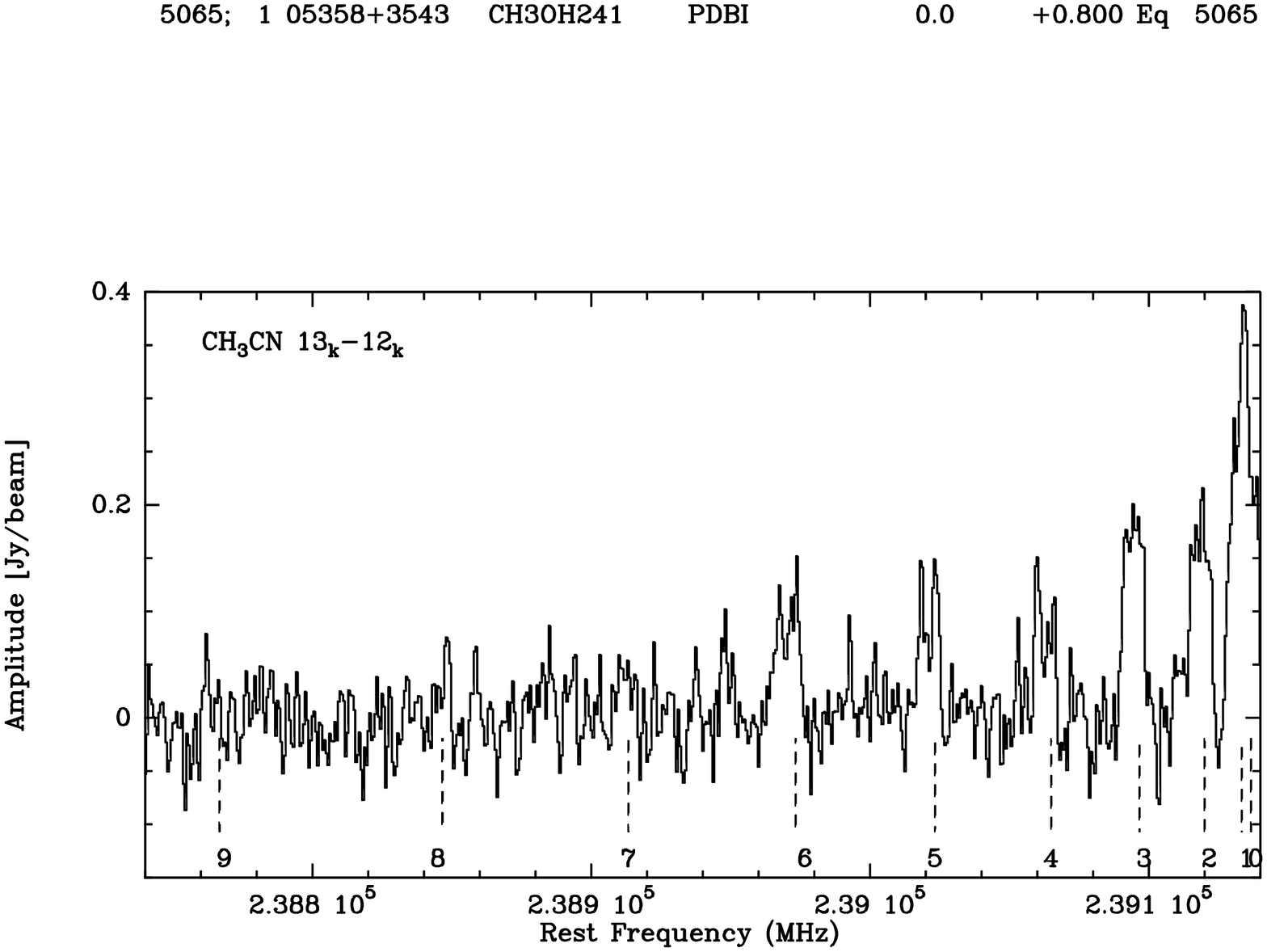}}
\subfigure[]{\label{1mm}
\includegraphics[bb= 200 16 552 681,clip, angle=-90,width=15cm]{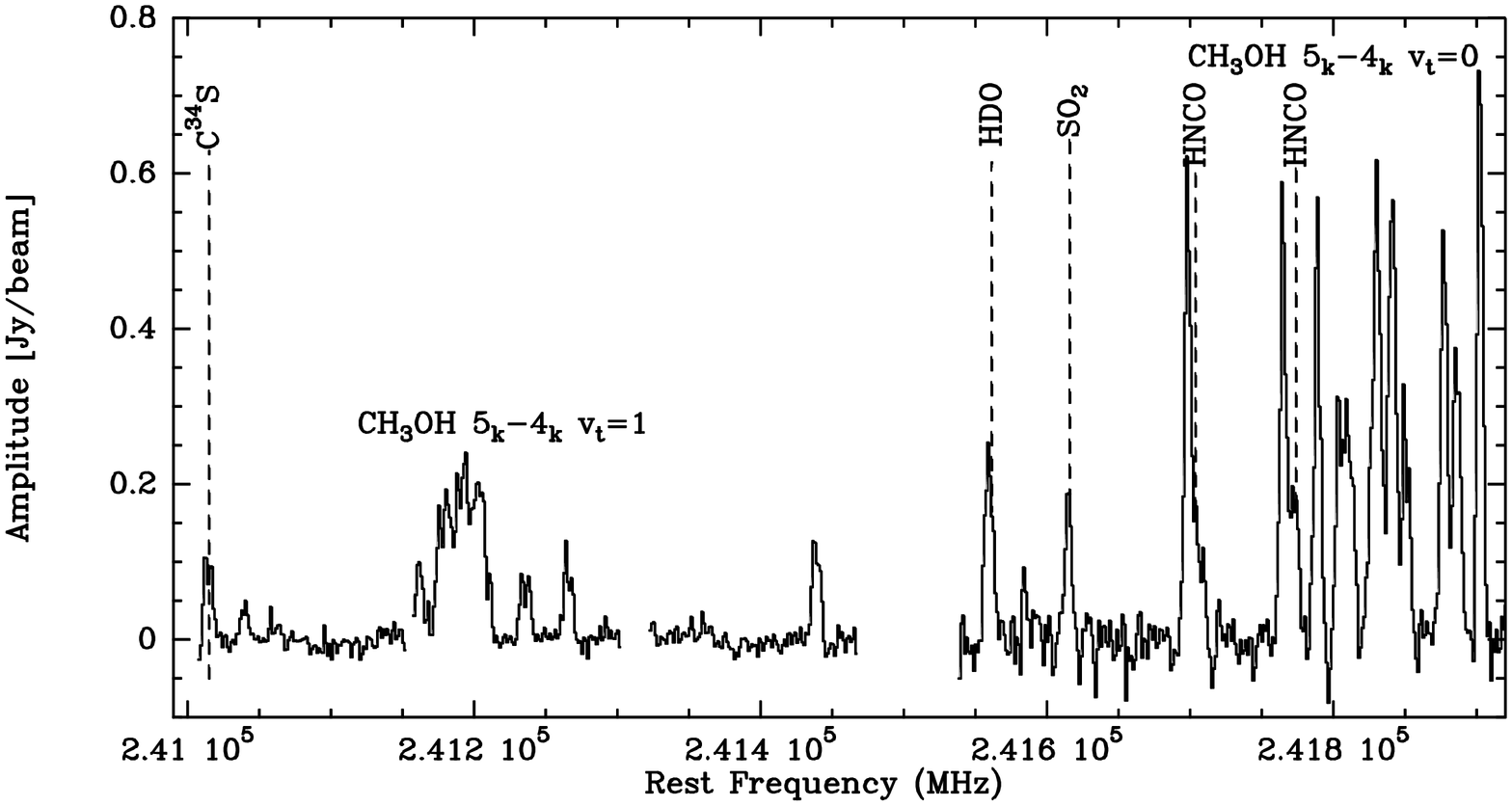}}
\subfigure[]{\label{338}
\includegraphics[bb= 200 16 552 681,clip, angle=-90,width=15cm]{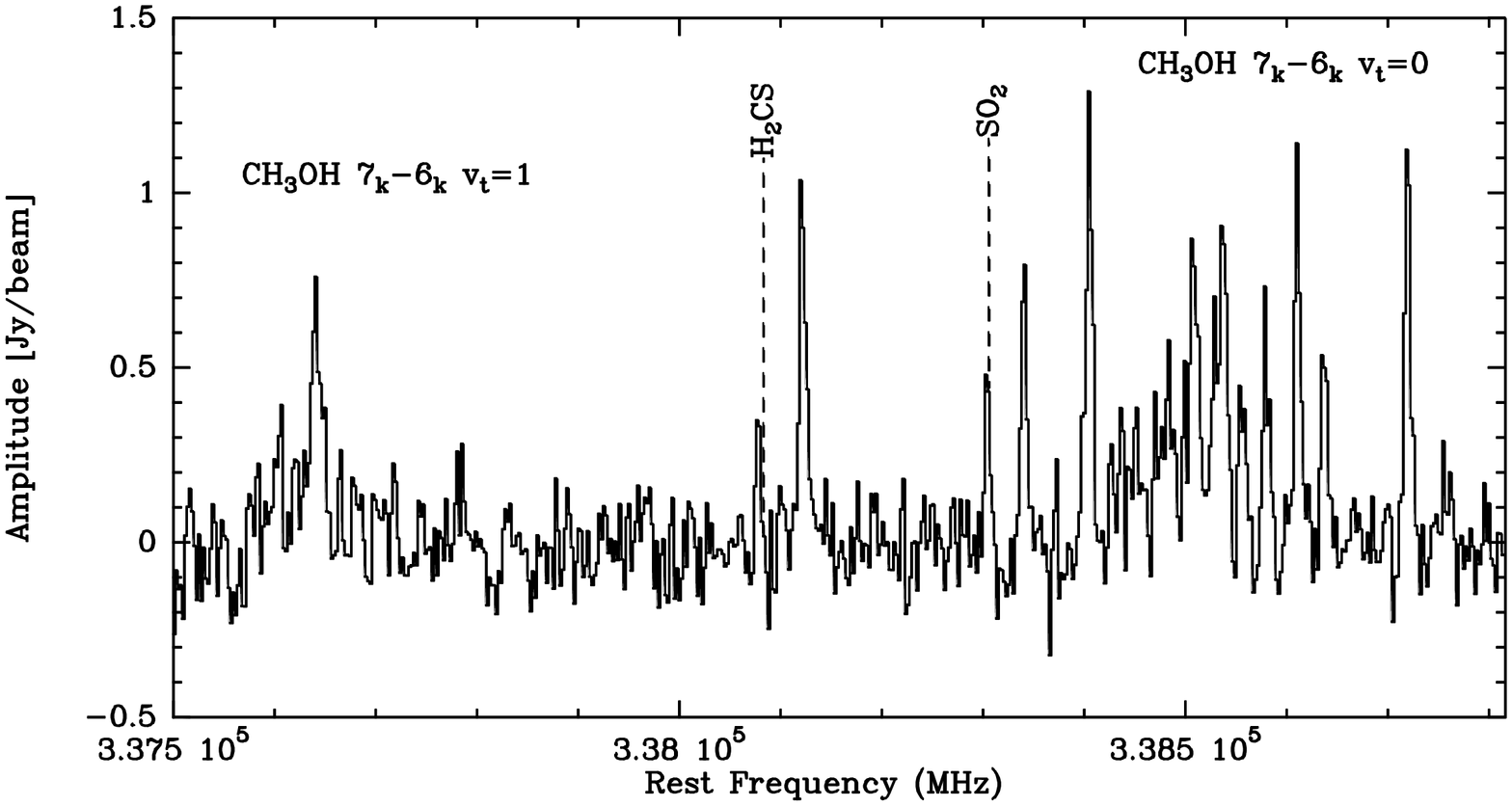}}
\caption{Molecular spectrum of mm1. The two cores, mm1a and mm1b, are resolved only for $5_k\to 4_k~v_t=1$ CH$_3$OH lines at $\sim 241.2$~GHz in Fig.\ref{1mm}.  For CH$_3$CN, different $K$ numbers are marked with dashed lines in the lower part of the spectrum (Fig.~\ref{ch3cn-mm1}). The features from other molecular species are labelled with the name of the molecule. Not labelled transitions are methanol lines.}\label{mm1-spectra}
\end{figure*}

{\bf Source mm1:} Figure~\ref{mm1-spectra} presents the observed
spectral bands at 238, 241 and 338~GHz, respectively, towards the main
dust condensation mm1. High excitation (E$_{up}>200$~K) transitions
are detected towards this core; however, the datasets where the two
dust condensations mm1a and mm1b are resolved, show that the emission
from high excited lines is associated only with mm1a. Only
the $5\to4$~C$^{34}$S transition ($v\sim -13.4$~km~s$^{-1}$) is detected on mm1b. The lack
of molecular emission from mm1b could be a bias of the observations:
the two dusts condensations are resolved only in the AB configuration
of the PdBI, which was aimed at the detection of torsionally excited
lines of methanol and had the $5\to4$~C$^{34}$S as the only low
excitation transition. The two condensations have very similar properties
for the continuum emission \citep{05358-beuther}.
However, other observations suggest that mm1a and mm1b are indeed
of different nature, as 
methanol class~II masers
\citep{2000A&A...362.1093M}, and a mid-infrared source
\citep{2006MNRAS.369.1196L} are detected on mm1a, but not on mm1b. A similar  chemical differentiation of two cores of very close masses was recently found by \citet{2007ApJ...658.1152Z} in the high-mass (proto)stellar cluster AFGL5142. As suggested for  AFGL5142, a possible interpretation of our observations is that mm1b is in an earlier evolutionary phase than mm1a, still not characterised by a rich chemistry. 

\begin{figure}
\centering
\includegraphics[bb= 184 17 559 763,clip, angle=-90,width=9cm]{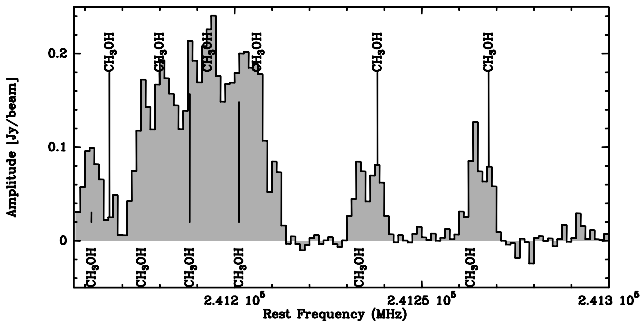}
\caption{Spectrum of mm1a at 241.2~GHz. The upper labels indicate
the rest frequencies of the CH$_3$OH lines for a velocity of
$-17.6$~km~s$^{-1}$, while the lower labels show the same transitions
for a velocity of $-11.6$~km~s$^{-1}$.}\label{components}
\end{figure}

{\it Source mm1a:} Several methanol transitions, in the ground state
as well as in the first torsionally excited level, show two velocity
components. The same profile is found in the $12_k\to 11_k$ CH$_3$CN
band (SMA data, Beuther priv. comm.) and in the $13_k\to 12_k$ lines, in
the $7_{1,6}\to 6_{1,5}$~H$_2$CS line, and in the $5\to 4$~C$^{34}$S
transition.  HNCO is  also showing the same behaviour in the 
 the $11_{0,11} \to 10_{0,10}$ line at $\sim -17.8$
and $\sim -15.3$~km~s$^{-1}$. Two peaks are also detected 
in the $11_{2,9} \to 10_{2,8}$
transition, but the blend with the $5_0\to 4_0$ CH$_3$OH-$E$ line and with 
the $11_{2,10} \to 10_{2,9}$ transition complicates the interpretation of this profile. On the other
hand, the SO$_2$ and HDO lines at $\sim 241.6$~GHz show only one
velocity component, peaked at -16.6 and -14.9~km~s$^{-1}$,
respectively. For the HDO line, this corresponds to the velocity in
between the two peaks detected in the other transitions.  An example of the double peak profile of mm1a is given in
Fig.~\ref{components}, where part of the $5_k \to 4_k, v_{\rm t}=1$
CH$_3$OH band is shown.  For the $7_{1,6}\to 6_{1,5}$ H$_2$CS and the
$5\to4$~C$^{34}$S lines, this profile could be caused by missing flux;
self absorption or missing flux problems could affect the low energy
transitions of methanol. The hypothesis of self absorption is
strengthened by the profile of  HDO transition. However, since the double-peaked profile is found also at
high energies, where the optical depth is unlikely to be high, even in
rare molecular species, we believe that the two peaks are, at
least for the torsionally excited lines of CH$_3$OH, for CH$_3$CN and HNCO,
real, and not due to self absorption.  In several cases, overlap
between different lines complicates the identification of the two
components; however, a few lines ($5_{0} \to 4_{0}$ and $5_{-1} \to
4_{-1}~v_t=1$ CH$_3$OH; $13_{4} \to 12_{4}$~CH$_3$CN) 
do not overlap with any other transitions. By Gauss
fitting their line profiles, we derive
linewidths of $\sim$3 and $\sim$4.~km~s$^{-1}$, and system velocities
of $\sim-17.6$~km~s$^{-1}$ and $\sim-11.6$~km~s$^{-1}$,
respectively. Both velocity components peak on mm1a. 

\begin{figure}[]
\centering
\subfigure[]{\label{dv}\includegraphics[angle=-90,width=8cm]{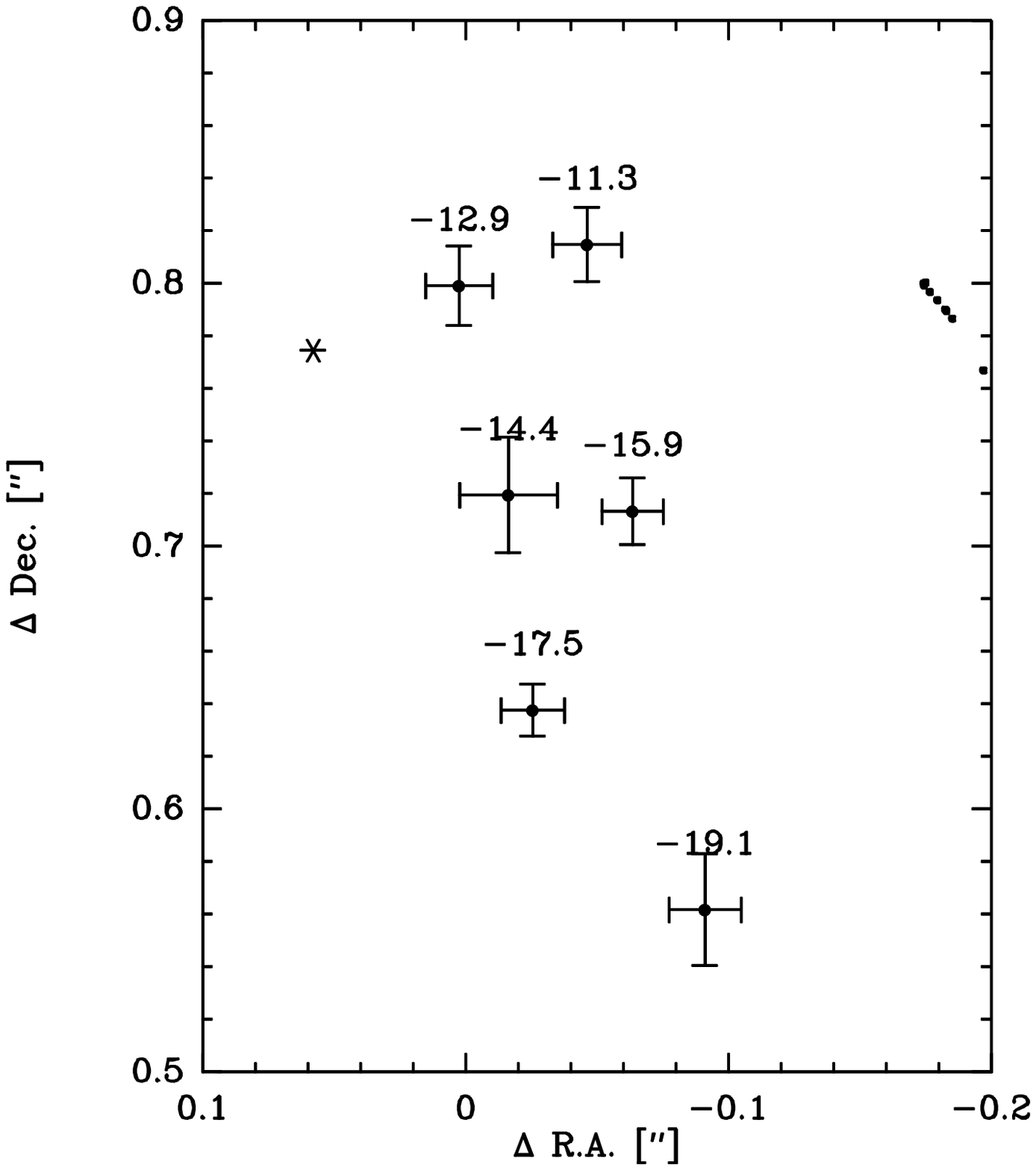}}
\subfigure[]{\label{ocs}\includegraphics[angle=-90,width=9cm]{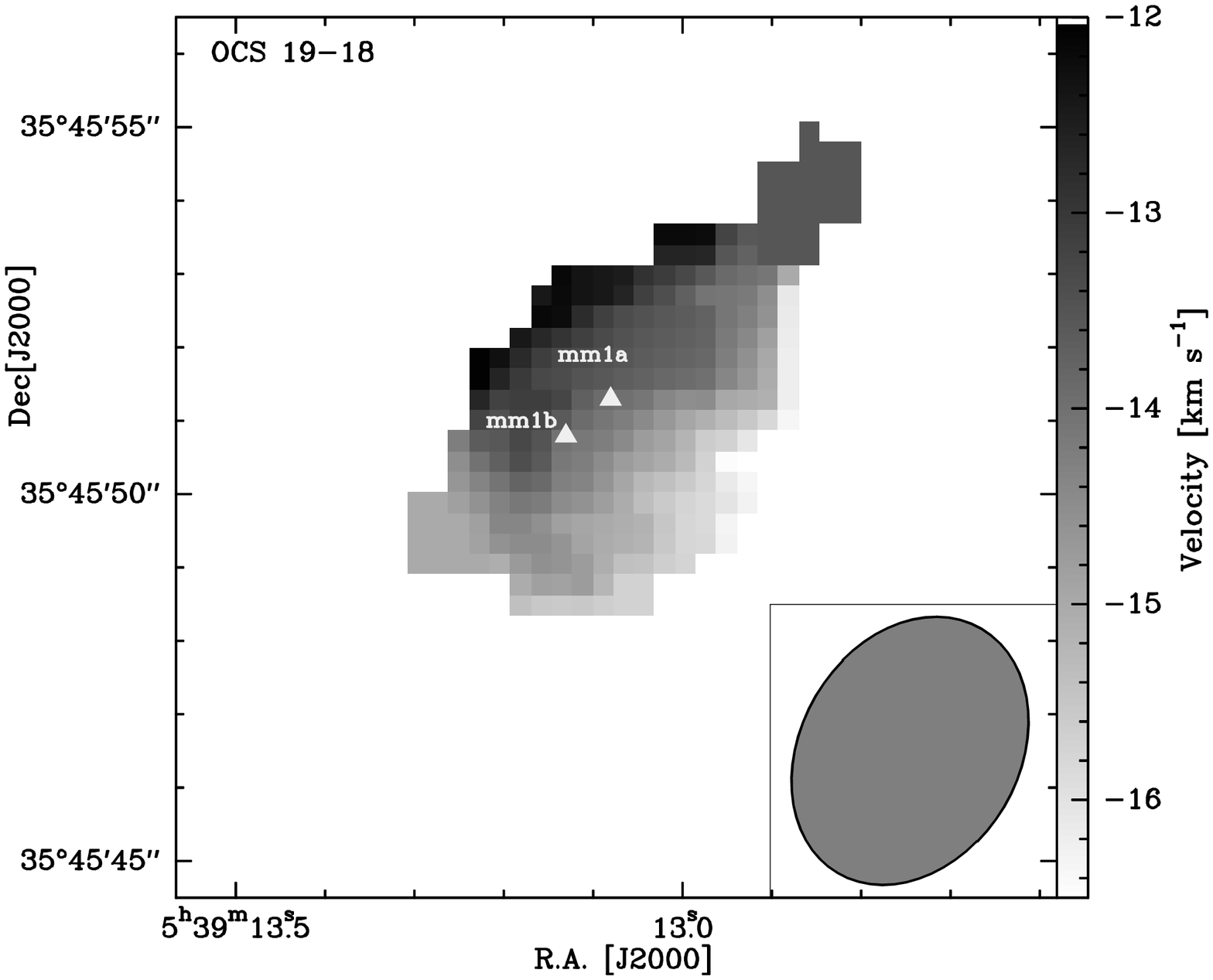}}

\caption{{\bf a:} Presented is the position-position (P-P) diagram of
the $5_{-1} \to 4_{-1}~v_t=1$-$E$ CH$_3$OH line.  The star marks the
position of the 1.3 mm continuum peak. The values above each mark are
the LSR velocities of the corresponding channels. The error bars
represent statistical errors of the fits. The error on the position of
the continuum peak are negligible. The black dots on the east mark the
positions of the velocity channels of the 6.7~GHz methanol maser
\citep{2000A&A...362.1093M}. The offset between the two linear
structures can be accounted for pointing uncertainties, and error in
the absolute position of the VLBI data. {\bf b:} First moment map of
the $19\to 18$~OCS transition. The white triangles mark the positions
of the dust condensations mm1a and mm1b. The beam is shown in the
bottom right corner.}\label{vel}
\end{figure}

We fitted the peak position of each velocity channels of the $5_{-1}
\to 4_{-1}~v_t=1$~CH$_3$OH-$E$ line, and detected a velocity gradient
along an elongated structure (Fig.~\ref{dv}). This linear structure
seems to be shifted from the millimetre continuum peak.  The same
velocity gradient of the $v_t=1$ methanol lines is found in the $19\to
18$~OCS line at 231~GHz (detected with the SMA, Beuther priv. comm.),
and in the 6.7~GHz methanol maser transition
\citep{2000A&A...362.1093M}. In Fig.~\ref{dv}, we show for comparison
the positions of the velocity channels of the maser line, and in
Fig.~\ref{ocs} the first moment map of the $19\to 18$~OCS
transition. Moreover, observations with the SMA with an angular
resolution of $\sim 3''$ (Beuther, priv. comm) show that the $2\to
1$~C$^{18}$O line is associated with the high velocity outflow of
\citet{2002A&A...387..931B}, and has a velocity gradient perpendicular
to the linear structure detected in methanol.  While an
unresolved expanding or contracting shell of gas would also produce
the double-peaked profile detected towards mm1a, the linear velocity gradient detected in methanol
and OCS cannot be explained by these scenarios.

The line profile detected towards mm1a is reminiscent of the
double-peaked profiles found in low and intermediate mass tilted
circumstellar disks, in optically thin and thick lines
\citep{1993ApJ...402..280B}.  Moreover, the double-peaked profile
arises from a region ($R\le 550$~AU) comparable in size to the
candidate disks surrounding high-mass (proto)stars \citep[for a
compilation of them, see][]{2007prpl.conf..197C}.  The detection of a
velocity gradient is not conclusive of a circumstellar disk. However,
we believe that the velocity structure detected in methanol and
perpendicular to the high-velocity outflow, together with the
double-peaked profile, and the size of the emitting gas are strongly
suggestive of a  rotating structure in
IRAS\,05358-mm1a. To investigate whether this structure is a circumstellar disk or a toroid like 
the ones discussed by \citet{2007prpl.conf..197C}, higher linear and velocity resolutions are needed 
to study the velocity field of the innermost gas around mm1a. We note however that toroids have 
typical radii of several thousands AU, while the radius of the structure around mm1a has an upper limit of  550~AU.

Alternatively, 
two different
hot components with a separation of less than $1100$~AU would also produce a double-peaked profile, and a velocity gradient along a linear structure.  Since the
two lines have very similar intensities and linewidths, the physical
conditions in the two cores would be comparable (as derived in
paragraph \S\ref{ch3cn}).
The total mass of the binaries can be derived via \citep[e.g.,][]{2006ApJ...639..975C}
\begin{equation}
M_{tot}=\frac{A\times \Delta v^2}{G\times sin^2(i)}
\end{equation}
where $A$ is the semi-major axis of the elliptical orbit of the two
stars, $G$ the gravitational constant, $i$ the inclination angle,
which is unknown, and $\Delta v$ the relative velocity between the two cores. An upper limit to the major axis is given by the
resolution of the data ($0.6''$, corresponding to $\sim 1100$~AU, at
this distance). In section~\S\ref{ch3cn}, the sizes of the
two sources are derived by fitting the $13_{k} \to
12_{k}$ CH$_3$CN band.  Therefore, the lower limit to the major axis of the
orbit is taken as the sum of their radii, corresponding to $0.2''$
(360~AU), which is in agreement with the extension of the linear
structure in Fig.~\ref{dv}. Hence, the total mass of the system ranges
between 22 and 7~M$_\odot$/sin$^2i$, in good agreement with the value of 13~M$_\odot$ derived from the Lyman continuum flux \citep{05358-beuther}.

In both cases, a rotating structure or two close-by hot cores, the single-peaked 
line profile of HDO would suggest that the emission is coming from a larger  envelope of gas.

To discriminate between the different possibilities discussed
above, higher spatial resolution observations will be required to
resolve and image the system in more
details. Among the current cm-mm interferometers, only the VLA in its
A configuration would allow observations with a spatial resolution
significantly higher than the one of our current data. Otherwise,
studies of such systems at resolutions of $0.1''$ or less in spectral
lines other than NH$_3$ will be possible in the submm windows only 
when the Atacama Large Millimeter Array
(ALMA) will come on line 
in the
next few years.

{\bf Source mm2:} the molecular spectrum of mm2 shows emission from
several species, with moderately excited ($E_{up}\le 130$~K) methanol
lines. Linewidths are comparable to the one of the components in mm1, and emission is at
v=$-15.3$~km~s$^{-1}$. However, stronger emission is detected in
several methanol lines at (1.2$'',0.6''$) off of mm2
($\alpha_{2000}$=05$^h$39$^m$12$^s$.86,
$\delta_{2000}$=$+$35$^{\circ}$45$'$51$''$.9), at a position where no
continuum emission is detected (see Fig.~\ref{ch3oh}).  At the same position, the CH$_3$CN
$k=0,1,3$ lines are also detected, while the $7\to6$ C$^{34}$S and $10_{1,10} \to 9_{1,9}$ 
H$_2$CS transitions have a secondary emission peak
in the vicinity of this position.  Linewidths are of the order of
3.7~km~s$^{-1}$, at the same velocity of mm2.

\begin{figure*}[]
\centering
\subfigure[]{\label{ch3cn2}
\includegraphics[bb= 200 30  552 688,clip, angle=-90,width=15cm]{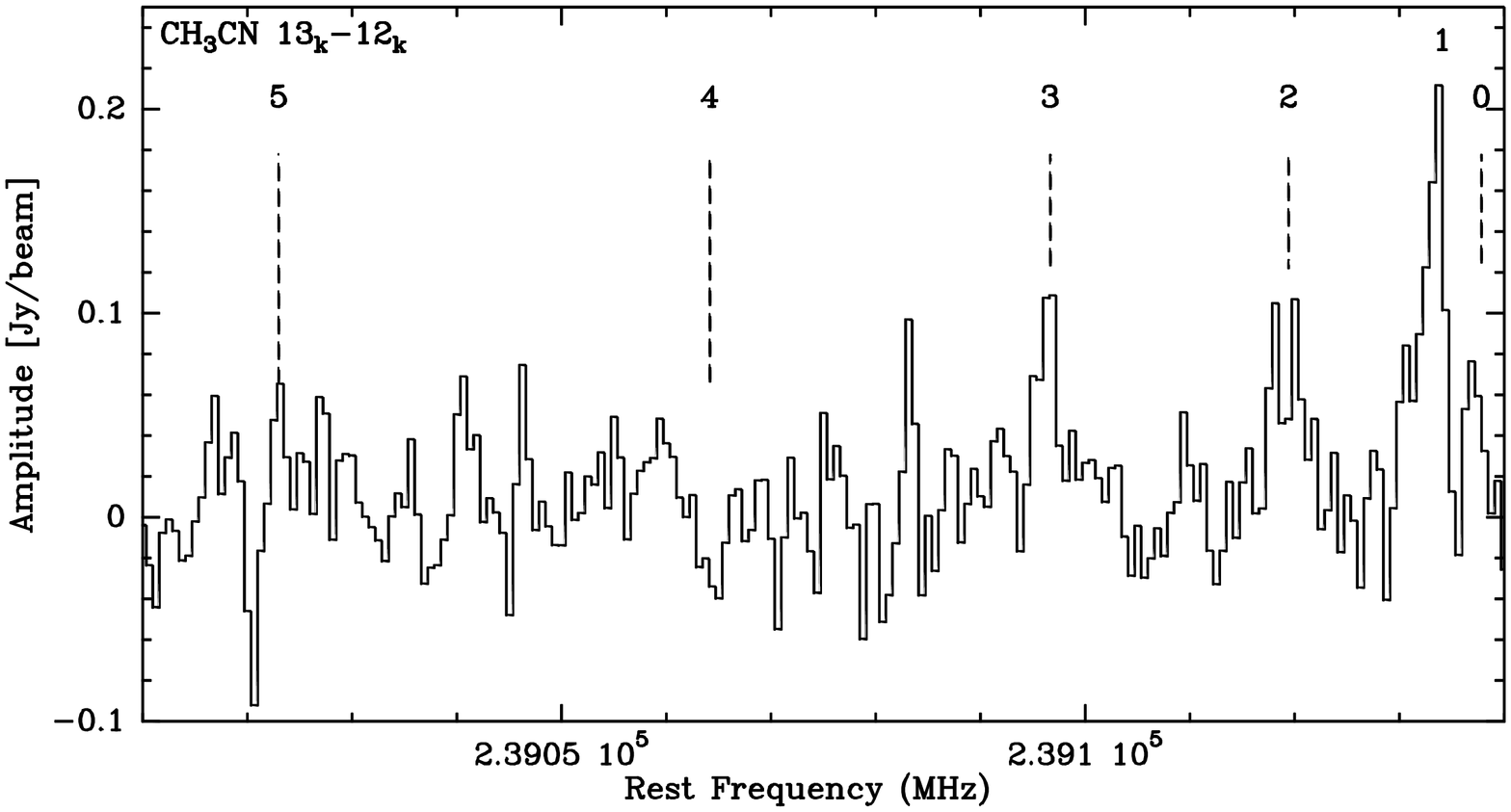}}
\subfigure[]{\label{ch3oh2}
\includegraphics[bb= 200 30  552 688,clip, angle=-90,width=15cm]{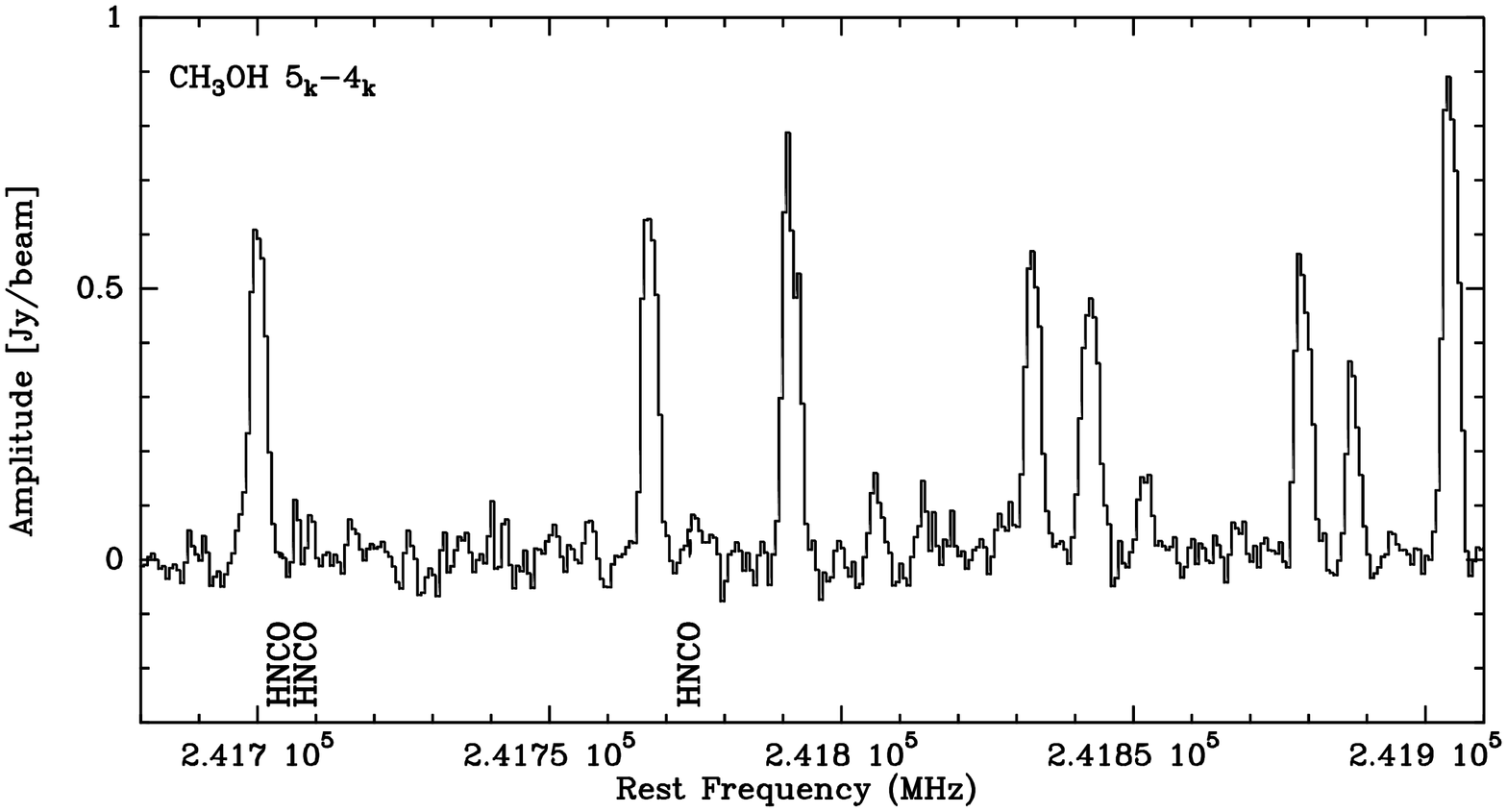}}
\subfigure[]{\label{3382}
\includegraphics[bb= 200 30  552 688,clip, angle=-90,width=15cm]{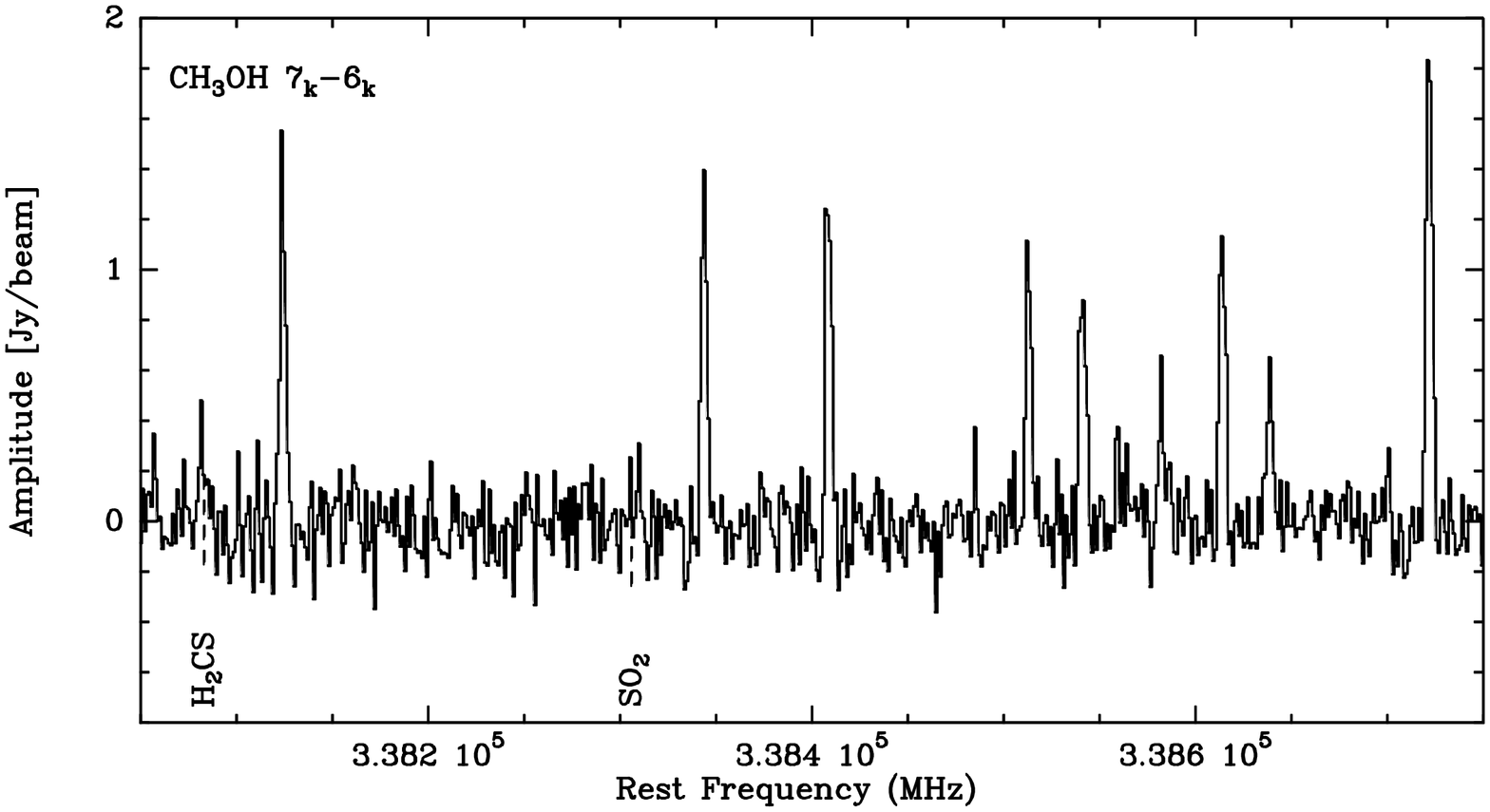}}
\caption{Molecular spectra towards mm2-line $[\alpha_{2000}$=05$^h$39$^m$12$^s$.86,
$\delta_{2000}$=$+$35$^{\circ}$45$'$51$''.9]$. For CH$_3$CN, different $K$ numbers are marked with dashed lines in the upper part of the spectrum (Fig.~\ref{ch3cn2}), while only the species are given for the other molecules. Not labelled transitions are methanol lines.}\label{mm2-spectra}
\end{figure*}

  To rule out the hypothesis that the different distribution of the
gas with respect to the dust is not real, but due to optical depth
effects, the intensity of the lines should be mapped in the wings,
where the optical depth is lower. Unfortunately, the signal to noise
in the CH$_3$CN spectrum and in the SMA data, which are the two
datasets less affected by line-overlapping problems, is not good
enough to perform this check. However, given the range of energies of
the lines showing a second peak between mm1 and mm2 (see
Fig~\ref{allspecies} and \ref{ch3oh}), we believe that this spatial
distribution is real. In Fig.~\ref{mm2-spectra}, the CH$_3$CN and
CH$_3$OH spectra are shown towards the offset position. In the rest
of the paper, we will refer to this position as to mm2-line. The
molecular spectrum of mm2 looks very similar to the one of mm2-line,
but line intensities are usually a factor of 1.2 weaker than on
mm2-line. Since the two positions have a separation of roughly half 
the beam size of our observations, we cannot rule out that part of the
emission at mm2 is coming from mm2-line.

\begin{figure}
\centering
\includegraphics[angle=-90,width=9cm]{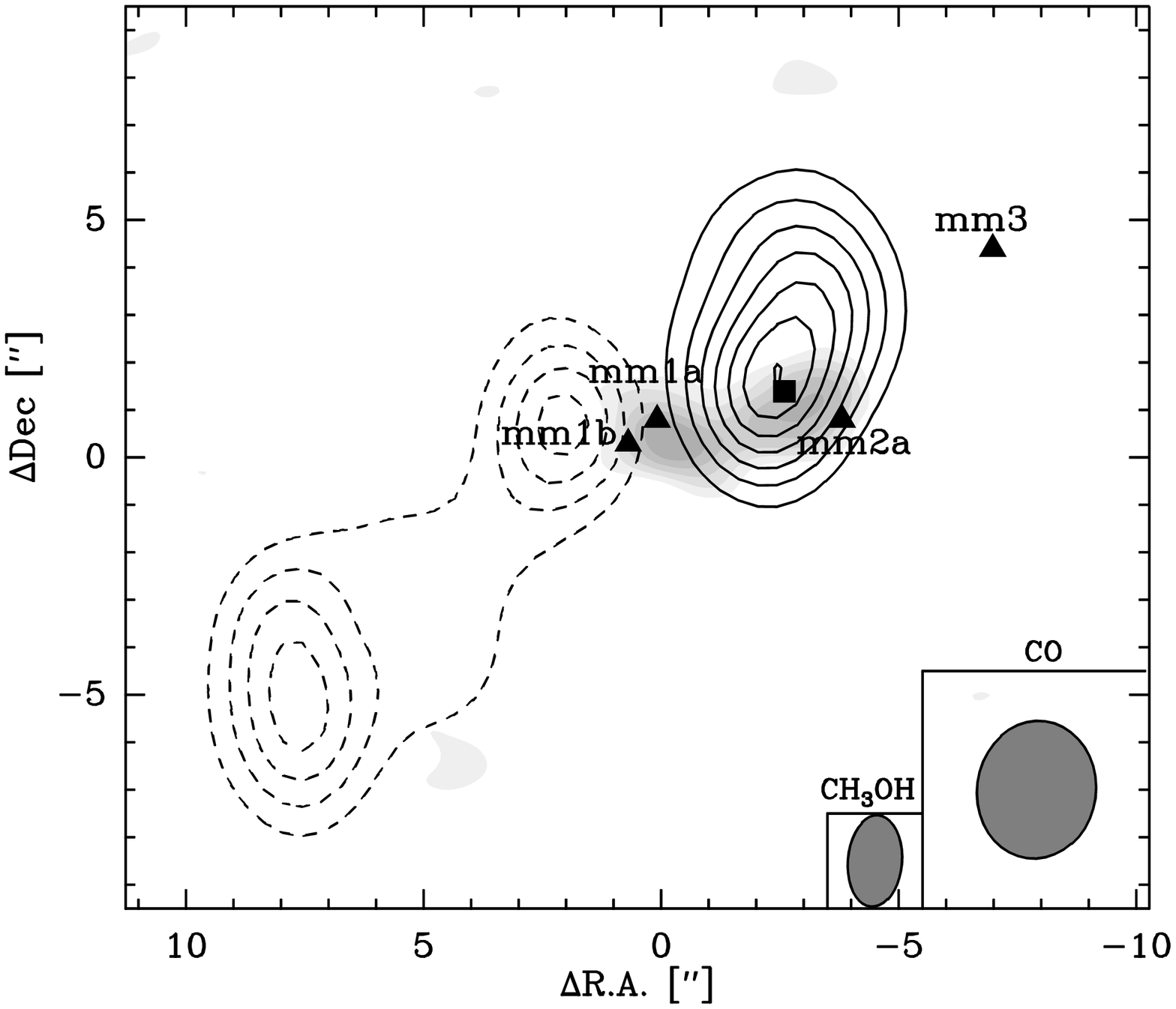}
\caption{In grey scale, the integrated intensity of the $7_2\to~6_2~\rm{v_t=0}$~CH$_3$OH-$E$ with the SMA (from 2.5~Jy~beam$^{-1}$~km~s$^{-1}$ in step of 1), overlaid on
the CO $2\to1$ high velocity outflow  (SMA, Beuther, priv. comm. Blue: solid lines $v=[-44,-24]$~km~s$^{-1}$; 
red: dashed lines $v=[-8,2]$~km~s$^{-1}$; levels from 1.5~Jy~beam$^{-1}$~km~s$^{-1}$ in steps of 1). The four triangles mark the
positions of the main mm dust condensations; the square outlines the mm2-line position.}\label{ch3oh-co}
\end{figure}

  A possible interpretation for the molecular
emission at mm2-line is that it is caused by sputtering or thermal
evaporation of the grain mantles, followed by gas-phase reactions, due
to the interaction with one or more of the outflows in the region (see
Fig.~\ref{ch3oh-co}).  Although mm2-line is characterised by
narrow features ($\Delta v_{LSR}\sim 3.2$~km~s$^{-1}$), the emission
is red-shifted respect to mm1, thus strengthening the outflow
scenario. However, the detection of a high-density tracing molecule
like CN$_3$CN at this position does not easily fit with this
hypothesis. Alternatively, a low-mass dense core could be responsible
for this emission. In this case, an upper limit to the mass of
mm2-line can be computed from the continuum emission at 1.3~mm,
assuming a temperature of 60~K as derived from the analysis of its methanol
spectrum (see \S~\ref{ch3oh-an}). This results in a mass of $M_{\rm{H_2}}\sim
0.08$~M$_\odot$. In \S~\ref{ch3oh-an}, we derive a lower limit of $10^{-7}$
for the abundance of methanol at mm2-line, which is comparable to the values found in 
low-mass star hot cores \citep{2005A&A...442..527M}. These cores are however usually warmer 
than 60~K. Otherise, the enhancement of the CH$_3$OH abundance can also be explained within  
the hypothesis that the
emission is caused by the interaction of molecular outflows with
the ambient molecular cloud.


{\bf Source mm3:} the core mm3 presents the simplest molecular
spectrum among the several dust condensations in the region. The
emission associated with mm3 is extended and not coming from a compact
core, as all the lines detected are seen as negative features, an
artifact due to missing spacings.  Therefore, only a qualitative
analysis of these data is possible.  \citet{05358-beuther} computed
its spectral energy distribution and derived an upper limit to the
dust temperature of $\sim20$~K, a value that is constrained by the weakness of the continuum emission at 690~GHz.  In paragraph
\S\ref{ch3oh-an}, we discuss the molecular spectrum at this position,
and conclude that the gas around mm3 is cold ($<30-40$~K), and dense
($10^5-10^6$~cm$^{-3}$).  Therefore, our results agree with
the conclusion of \citet{05358-beuther} that mm3 is a candidate of a
cold massive core in an early evolutionary stage. Our observations of CH$_3$OH and C$^{34}$S
suggest that these molecules have an extended, flat distribution, and
show no trace of the structure seen in the dust continuum emission, a
result reminiscent of observations of low-mass starless cores
\citep[e.g.,][]{1999ApJ...523L.165C}. Even H$^{13}$CO$^+$, a good
tracer of quiescent gas, shows no peak on mm3
\citep{2002A&A...387..931B}.  These results         suggest that mm3 is indeed a 
 candidate massive starless core.
However,  high angular resolution observations that trace the dense gas and
investigate potential signs of star formation activity are needed to test this hypothesis.

\section{Derivation of physical parameters}\label{analysis}
\subsection{CH$_3$CN}\label{ch3cn}
The physical properties of the gas around mm1 are derived by the
analysis of the 13$_k \to 12_k$~CH$_3$CN lines.  
The analysis was carried out with the Xclass program \citep[discussed in][]{2005ApJS..156..127C}, which uses 
an LTE
model to produce synthetic spectra, and compares them to the
observations. The 
molecular data are from the CDMS \citep{2001A&A...370L..49M} and JPL \citep{pickett_JMolSpectrosc_60_883_1998} 
databases. The parameters defining the synthetic spectrum are: source size, 
rotation temperature, column density, velocity width and
velocity offset (with respect to the systemic velocity of the object).
Several velocity components, which are supposed to be non-interacting
(i.e. the intensities add up linearly), can be used. 

Given the low signal-to-noise ratio of the spectra, 
it is  possible to perform this analysis only on mm1. For this position,
we assumed that all the CH$_3$CN transitions trace the same warm, dense
gas around the central object and come from the two velocity components
detected in several molecular species (see paragraph \S\ref{proto}), although the lower excitation lines
may have contribution from a more extended component (Fig.~\ref{allspecies}).
Since the line profiles
consist of two peaks of the same intensity, we used the same physical conditions for the two components. The source size is degenerate with temperature in the case of
completely optically thick lines, and with column density for
completely optically thin lines. In our case, the $k>3$ lines are
optically thin, while the $k\le3$ are optically thick. Therefore,
the degeneracy between source size and column density can be
solved. Constrains on the column density of CH$_3$CN come also from
the non detection of the $13_k\to 12_k$~CH$_3^{13}$CN band, which is
very close in frequency to the CH$_3$CN lines.  We found
that the source size is constrained to $\sim0.2''$ for each
components, in agreement with the size of 0.2-0.3$''$ of the linear
structure of Fig.~\ref{dv}.  For this value, the best fit corresponds
to a temperature of 220~K, and a column density of $\sim 4\times
10^{16}$~cm$^{-2}$ per velocity component. The uncertainties on these values are reported in Table~\ref{par}. Assuming that the CH$_3$CN
emission comes from mm1a, and using the hydrogen column density
derived by \citet{05358-beuther} for the AB configuration data at
1.2~mm ($N_{\rm H_2}\sim 2.2\times 10^{24}$~cm$^{-2}$), this results
in a total abundance of CH$_3$CN relative to H$_2$ of $4\times 10^{-8}$,
which is typical of hot cores \citep[e.g.,][]{1998A&AS..133...29H}.
However, \citet{05358-beuther} derived the H$_2$ column density for a
temperature of 50~K, while our estimate of $T_K$ is higher. The
abundance increases to $ 2\times 10^{-7}$, when re-computing the H$_2$
column density of mm1a at 220~K ($N_{\rm H_2}\sim 4.6\times
10^{23}$~cm$^{-2}$).

\begin{figure}
\centering
\includegraphics[angle=-90,width=9cm]{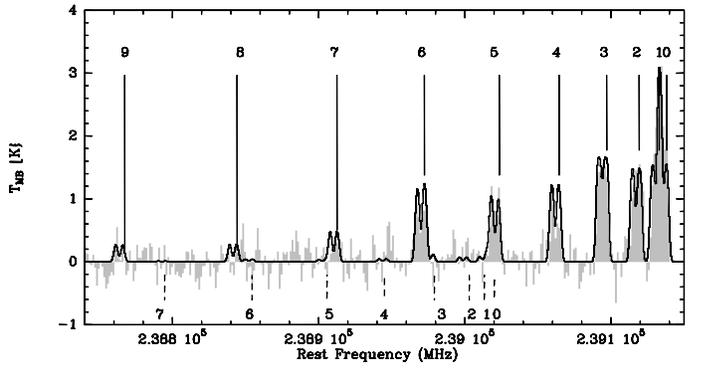}
\caption{Spectrum of the 13$_k \to 12_k$~CH$_3$CN band towards the
main dust condensation mm1; overlaid on the data, in black, the best
fit synthetic spectrum. All CH$_3$CN lines are labelled in the upper axis for the velocity
component $v=-17.6$ km s$^{-1}$. The frequencies of the corresponding CH$_3^{13}$CN transitions are labelled in
the lower axis.}\label{ch3cn-fit}
\end{figure}
\subsection{SO$_2$}\label{so2-an}
Four SO$_2$ transitions were detected towards the main condensation
mm1. They range in energy of the upper level from $\sim 24$ to 200~K; therefore, although they look
point-like in all observations,  they likely trace different gas around mm1. 
 We analysed the SO$_2$ emission by means of the LTE technique
discussed in paragraph \S\ref{ch3cn}. Since the lines do not show any trace of the double-peaked profile detected in other molecular species, we
used a model with one velocity component for SO$_2$.
Assuming a source size of 1$''$, the best fit is found for $T=150$~K and $N($SO$_2)=10^{16}$~cm$^{-2}$,
which corresponds to an abundance of SO$_2$ relative to molecular
hydrogen of $2\times 10^{-8}$ using $T_{dust}=50$~K ($N_{\rm H_2}\sim 6.6\times 10^{23}$~cm$^{-2}$), and of $\sim 5\times 10^{-8}$ for a temperature of 150~K ($N_{\rm H_2}\sim 2\times 10^{23}$~cm$^{-2}$). The errors on the parameters are listed in Table~\ref{par}.

\subsection{CH$_3$OH}\label{ch3oh-an}

For the analysis of the CH$_3$OH spectrum, we used a technique similar
to the one discussed for CH$_3$CN (\S\ref{ch3cn}) and implemented to the
LVG approximation \citep{2004A&A...422..573L,2007A&A...466..215L}. Since methanol is detected
at several positions, in the following discussion the properties of each dust condensation
will be discussed separately. The errors on the parameters are listed in Table~\ref{par} 
for each condensation.

{\it Source mm1a:} In our analysis of mm1a, we included only emission
coming from the $v_t=1$ lines, as their optical depth is lower than
for the ground state, and their emission is confined to the gas around
the dust condensation, while the $v_t=0$ transitions are more
extended, and are affected by problems of missing flux. Moreover, the
$v_t=1$ lines have similar level energies, while the ground state
transitions in our data range from a few K to $\sim 200$~K, thus not
tracing the same gas.  
The model includes the three torsionally excited bands
observed, the $2_k \to 1_k$ at 3~mm, the $5_k \to 4_k$ at 3~mm, and
the $7_k \to 6_k$ at 0.8~mm.   
Infrared pumping by
thermal heating of the dust is included in the form of an external
grey-body \citep{2007A&A...466..215L}, with $\beta=1.6$ and
$\tau_{100\mu m}$=1.5. This is necessary to explain the excitation of
the torsionally excited lines, as, with very high critical densities
($ 10^{10}$--10$^{11}$ cm$^{-3}$) and high level energies ($T\ge
200$~K), they can be hardly populated by collisions, but trace the IR
field instead. On the other hand, the ground state lines can be populated by collisions and
by the infrared pumping. 
\citet{2007A&A...466..215L}
investigated the effects of infrared pumping on the excitation of
methanol, and found that, as in the case of other molecules
\citep[e.g. CS, ][]{1981ApJ...245..891C}, the infrared pumping mimics the
excitation by collisions of the $v_t=0$ lines. This results in a
degeneracy between the density  and  the infrared field
of the thermal dust which makes any determination of the density through the analysis of the ground state lines of methanol impossible, for sources like hot cores. As in the discussion in paragraph
\S\ref{ch3cn}, we used two velocity components to model the data.

  The best fit is reached for values very similar to the
ones derived with CH$_3$CN, source sizes of 0.15$''$ and a temperature
of 220~K. As the line intensities are very similar in the two velocity
components in almost all lines, the same parameters are derived for
both of them. The column density for methanol is equal to $\sim
2\times 10^{18}$~cm$^{-2}$ per component, which, using the H$_2$
column density derived by \citet{05358-beuther} for mm1a for 50~K,
corresponds to a total abundance of methanol relative to molecular
hydrogen of $2\times10^{-6}$, or to $10^{-5}$ if we adopt 220~K as
dust temperature. Both values are typical of massive hot cores
\citep{1986A&A...157..318M,1988A&A...198..253M}.

The
fit well reproduces the intensities of the $J=5,7$ series, but it
fails to correctly predict the strength of the lines at 96~GHz,
which are weaker in the model than in the observations.  A plausible
explanation is that 3~mm lines come from a more extended region than
the others, and therefore the real beam dilution is less than that
obtained in the model with a source size of 0.15$''$. With the
resolution of our data, we cannot verify this hypothesis. The $2_k\to
1_k$ lines were observed in both the BCD and the AB configurations of
the array, and no flux is lost in the data with the higher
resolution. However, the CLEAN beam of the AB data at 3~mm is
significantly larger than the one at 1.3~mm, and different emission
sizes between the 3~mm and the 1.3~mm lines cannot be ruled out. 
Moreover, the 96~GHz lines peak at 300~K in
energy of the upper level, while the other bands are at higher
energies (330--530~K), and they are therefore more likely to trace the
gas closer to the central heating source.

{\it Source mm2:} Since no torsionally excited lines are detected toward
mm2, the analysis on the methanol spectrum was performed on the ground
state lines, and no external radiation field was used in the
calculations.  As discussed for mm1a, the analysis of these lines is
not trivial, as they range in energy of the upper level from 35~K to
more than 100~K (see Table~\ref{ch3oh-lines}), and therefore trace
different regions of gas. Moreover, some $v_t=0$ lines have extended emission and are,
thus, affected by missing fluxes problems.  The analysis of the
datacubes at 241.7 and 338~GHz shows that several methanol lines have
a spatial distribution similar to that of the $5_0\to 4_0~v_t=0$~CH$_3$OH-$A$ transition
(Fig.~1(b)). We excluded these lines from our study of mm2, and
limited the analysis to the transitions that are peaked on the main
dust cores. An asterisk in Table~\ref{ch3oh-lines} marks the lines
with extended emission, excluded from the analysis of mm2.

The continuum data \citep{05358-beuther} reveal a complex area around
mm2, with several sub-sources which may not be of protostellar nature,
but which could be caused by the outflows in the region. Given the
relatively low resolution of the observations of the methanol ground
state lines, the molecular emission detected toward mm2 could arise
not only from the gas associated with the dust core mm2a, but
also from the other sub-sources around it.  Therefore, in modelling
the molecular spectrum of mm2, we used the beam size as source size,
and derived the average physical parameters of the gas over the
beam. This approach is justified by the low optical depth of the
methanol lines for the physical conditions found in the analysis. 
The same technique  was used to model the methanol
emission at the position ($\alpha_{2000}$=05$^h$39$^m$12$^s$.86,
$\delta_{2000}$=$+$35$^{\circ}$45$'$51$''$.9), north-west of
mm2. Since no continuum emission is detected there, the nature of the
molecular emission is unknown. The spectra toward the two positions, mm2
and mm2-line, are very similar: emission comes from the same
transitions, which have similar linewidths, and emit at similar
velocities.  However, the emission from mm2-line is stronger than on
mm2 in all methanol lines, hence suggesting that the methanol abundance is
enhanced at this position. 
The best fit is found for a temperature of 60~K,
and a column density of $4 \times 10^5$~cm$^{-2}$ and $7 \times 10^5$~cm$^{-2}$ for mm2 and mm2-line, respectively.
The column densities at the two positions are $4\times$10$^{15}$~cm$^{-2}$ for mm2, and 
$7\times$10$^{15}$~cm$^{-2}$ for mm2-line.
 The corresponding abundances of methanol relative to
H$_2$ can be calculated by using the H$_2$ column density derived by
\citet{05358-beuther} from the BCD configuration data. The methanol abundance of mm2
corresponds to $6\times10^{-9}$, a value closer to the ones derived
towards the inner regions of low-mass protostars
\citep[e.g.,][]{2005A&A...442..527M}, than to values found in
high-mass protostars
\citep{1986A&A...157..318M,1988A&A...198..253M}. In
\S~\ref{proto}, we noticed that, given the separation between mm2 and
mm2-line, at least part of the emission of mm2 could come from the
offset position. Therefore, the CH$_3$OH column density derived for
mm2 should be regarded as an upper limit to its true value. 

An upper limit to H$_2$ column density of mm2-line can be computed  from the continuum emission at 1.3~mm, $N\sim 5\times 10^{22}$~cm$^{-2}$ at 60~K.
This results in an abundance of $1\times 10^{-7}$ for methanol, which is  a factor of 17 higher than on mm2. However, a different dust temperature
would result in different values for $N_{H_2}$ and $X_{\rm CH_3OH}$. The
enhancement of methanol is usually ascribed to evaporation of icy
grain mantles in the vicinity of warm embedded sources, e.g. hot cores,
or to desorption of grain mantles in shocks associated to molecular
outflows \citep[e.g.,][]{1995A&A...295L..51B}. Since no embedded
sources are detected in the continuum emission at this position down to an upper limit of $0.08$~M$_\odot$, it
seems likely that the molecular emission associated with mm2-line is
related to the outflow activity of the region.

 In the case of mm2, and mm2-line, the temperatures derived from methanol are comparable to those
 adopted by \citet{05358-beuther} for the analysis of the continuum emission; therefore, only one value
 for the methanol abundance is given at this  position.

{\it Source mm3:} As listed in Table~\ref{ch3oh-lines}, only molecular
transitions from the ground state, and with upper energies of less than 70~K are detected toward mm3.
Since all the lines detected toward this core are seen in absorption due to missing fluxes, with
emission only in the $5_0\to4_0$-$A$ and $5_{-1}\to4_{-1}$-$E$ transitions
on top of the negative features, only a qualitative analysis of the
data is possible.  We ran LVG models for several temperatures and
densities, to find the range of parameters compatible with our
observations.  The detection of the $5_{\pm2}\to 4_{\pm2}$-$E$ lines,
and the non detection of the higher $k$ transitions constrain the
H$_2$ density to the range $\sim 10^5$--$ 10^6$~cm$^{-3}$, while the
tentative detection of the $k=0$-$A$ and $k=-1$-$E$ lines in the
$7_k\to6_k$ band, but not of the other transitions, sets an upper
limit to the temperature of 30$-$40~K.

\subsection{Uncertainties on the physical parameters of the gas}\label{errors}
\begin{table*}
\centering
\caption{Overview of the physical parameters in IRAS\,05358+3543: $-$ indicates 
parameters which cannot be derived with our data.}\label{par}
\begin{tabular}{lcccccc}
\multicolumn{1}{c}{Core} &\multicolumn{1}{c}{$\Theta$}&\multicolumn{1}{c}{T}&\multicolumn{1}{c}{n}&\multicolumn{1}{c}{$N$}&\multicolumn{1}{c}{$X$}&\multicolumn{1}{c}{$N_{\rm H_2}^a$}\\
&\multicolumn{1}{c}{$['']$}&\multicolumn{1}{c}{[K]}&\multicolumn{1}{c}{[cm$^{-3}$]}&\multicolumn{1}{c}{[cm$^{-2}$]}&\multicolumn{1}{c}{$$}&\multicolumn{1}{c}{[cm$^{-2}]$}\\
\hline
\hline
\multicolumn{7}{c}{CH$_3$OH}\\
\hline
mm1a$^b$&0.15&$220~(75-270)$&$-$&$4\times 10^{18}$~$(4\times 10^{17}-1\times 10^{19})$&$10^{-5}$&$4.6\times 10^{23}$\\
mm2&2.5&60~(40--80)&$>6\times 10^6$&$4\times 10^{15}~(4\times 10^{15}-5\times 10^{15})$&$6\times 10^{-9}$&$2.7\times 10^{23}$ \\
mm2-line&2.5&60~(40--110)&$>6\times 10^6$&$7\times 10^{15}~(1\times 10^{15}-2\times 10^{16})$&$10^{-7}$&$5\times 10^{22}$\\
mm3&$-$&$<40$&$10^5-10^6$&--&--&$4.9\times 10^{24}$\\
\hline
\multicolumn{7}{c}{CH$_3$CN}\\
\hline
mm1a$^b$&0.2&220~(120$-$330)&--&$8\times 10^{16}~(4\times 10^{16}-4\times 10^{18})$&$2\times 10^{-7}$&$4.6\times 10^{23}$\\
\hline
\multicolumn{5}{c}{SO$_2$}\\
\hline
mm1&1.0&150~($>60$)&$-$&$~1\times 10^{16}~(5\times 10^{15}-5\times 10^{16})$&$5\times 10^{-8}$&$2\times 10^{23}$\\\\
\hline
\hline

\end{tabular}
\begin{list}{}{}
\item $^a$ derived from the 1.3~mm continuum data presented by \citet{05358-beuther} and using the temperature derived from each molecular species
\item $^b$ the values refer to the total column density of the two velocity components.
\end{list}
\end{table*}
Error ranges for the various parameters can be estimated by a $\chi^2$
analysis, although this can only give errors within the assumptions of
the model. It cannot assess errors due to the LTE and LVG assumptions and to
the treatment of the gas as consisting of a finite number of
non-interacting components of homogeneous conditions.  As for the LTE
assumption, densities in hot cores usually are in a range which makes
it reasonable. 

The analysis of CH$_3$CN and CH$_3$OH infers a source size of
$\sim0.2''$ for the gas responsible of the emission of the two
velocity components found toward mm1a. This value is relatively well constrained
by the CH$_3$CN band, which shows optically thick and thin
lines, and by the non detection of CH$_3^{13}$CN. Since this result is in agreement with the size of the linear
structure found from the fit of the positions of the velocity channels
of methanol (see Fig.~\ref{dv}), we fixed the source size to this value, and derived
confidence levels for the estimates of the temperatures and column
densities.  However, one should remember that for optical thin
lines, like the $v_t=1$ CH$_3$OH transitions, the source size and column
density are degenerate parameters, and the estimates of the column
densities would change, if the source size were different than our
assumption.   For CH$_3$OH toward mm1a, since these data were taken with a beam of $\sim
0.7''$ and we assumed a size of $0.15''$ in the calculations, the
CH$_3$OH column density can only be a factor of 10 smaller than indicated by our
results. For a source size of 0.15\arcsec, good
fits to the data are found also for lower temperatures, but higher
column densities. For $T\sim100$~K the model requires
column densities of $\ge 10^{19}$~cm$^{-2}$, which would correspond to
very high abundances relative to H$_2$ for such temperatures  ($X_{\rm CH_3OH}\ge 10^{-5}$).
For
temperatures higher than 150~K, the methanol column density is
confined to the interval $4\times 10^{17}-2\times 10^{18}$~cm$^{-2}$. A lower limit to the kinetic temperature
is found at $\sim 75$~K, where
the torsionally excited lines of methanol are not anymore efficiently pumped.

Similarly, the column density of SO$_2$ on mm1, and of CH$_3$OH on mm2
and mm2-line would increase if they were constrained to a region smaller in size
than the value used in the calculation. For example, the column density of SO$_2$ 
would increase of one order of magnitude
by using a sorce size of 0.2$''$ as derived from the analysis of CH$_3$CN and CH$_3$OH. 

The $3~\sigma$~confidence levels derived for each parameter from the $\chi^2$ analysis are given in Table~\ref{par}, together with their best fit
values.

\section{Various stages of star formation}
The analysis of the spectral line emission in IRAS\,05358 reveals four
 dust condensations with a maximum projected distance of 18\,000~AU. At least three of them show signs of active star formation.
The
core mm1a appears to be the most evolved source of the region,
associated with a hypercompact H{\sc ii} region, and class II methanol
masers. \citet{05358-beuther} derived the Lyman continuum flux of mm1a
corresponding to a B1 ZAMS star of 13~M$_\odot$, {and a large reservoir of accreting material ($\sim 10$~M$_\odot$). 
However, this estimate is obtained from relatively low resolution data, where the cores mm1a and mm1b are not spatially resolved. From high resolution data, which are however affected by missing fluxes, the estimate to the mass of mm1a is 1~M$_\odot$, and 0.6~M$_\odot$ for mm1b.
Our observations suggest that mm1a also harbours a candidate
circumstellar disk.  A rough estimate of the luminosity of mm1a can be derived through the
Stefan-Boltzmann formula, assuming a radius of 180~AU, and a
temperature of 220~K. By adding the values for the two components, the total
luminosity of mm1a is $L\sim 6000~L_\odot$, in agreement with the luminosity of $10^{3.72}~L_\odot$ of a B1 ZAMS star
\citep{1973AJ.....78..929P}. Since the total luminosity of the cluster is $6300~L_\odot$, this result supports the interpretation 
that mm1a is the main powerhouse of the region.
 
\citet{2004MNRAS.354.1141V} derived the abundances of several
molecules in the gas phase, as function of the age, and of the mass
of the heating central object. The abundances of CH$_3$OH,
CH$_3$CN, and SO$_2$ of mm1a are in agreement with  a 15~M$_\odot$ star of approximately $10^{4.6}$~yr, 
although the best fit 
value for the column density of CH$_3$OH is higher than the theoretical values, but still consistent with the models 
 within the uncertainties of the fit. 
Decreasing the mass of the heating source
in the models results in increasing its
age. The value of $10^{4.6}$~yr is in very good agreement with the
value obtained by \citet{2002A&A...387..931B} for the dynamical timescale of the molecular outflows originating 
from the vicinity of mm1 ($t\sim 37\,000$~yr).

 The physical parameters of the gas derived from
our analysis, (temperature and abundances) are similar to hot cores,
thus implying that mm1a has already reached this evolutionary
stage. However, its molecular spectrum differs from the one of typical
massive hot cores.  \citet{beuther-g29} studied the hot core in
G29.96-0.02 \citep[$L\sim 9\times 10^4$~L$_\odot$,][]{2003A&A...407..225O} with the SMA,
with the same frequency setup we used for our SMA observations.
Although the linear resolution reached for G29.96 is comparable to
ours, the spectrum of G29.96 shows stronger
methanol lines than IRAS\,05358+3543 and higher excitation transitions; moreover, molecular species which
are detected in G29.96 are not in our dataset.  The same is found when comparing our observations 
with Orion-KL $(L\sim
10^5$~L$_\odot$), even with the poor angular resolution of the CSO
telescope \citep{1997ApJS..108..301S}.  Possible explanations for these
differences are that mm1a is indeed {\it a)} in an earlier
evolutionary stage than the other two sources, at the beginning of the
hot core phase, and will eventually reach their  chemical
richness; or {\it b)} hosts a less luminous hot core,
which will never produce the rich chemistry of G29.96-0.02 or
Orion-KL.  Both interpretations render mm1a an interesting source,
since they collocate it in a different region of the age-luminosity
plane of hot cores, which still has not received much
attention. Studies of large samples of hot cores at the same linear
resolution,  and
with the same frequency setups \citep[similar to that of][ but on a larger sample of sources]{2007IAUS..237..148B} 
are necessary to better understand
this evolutionary phase of star formation. Particularly interesting is
the comparison with observations of hot cores in the vicinity of low-
and intermediate-mass protostars, which could clarify whether the
poorness of the chemistry in IRAS\,05358+3543 is due to its age or it
is rather an intrinsic property of less luminous hot cores. 

Although the mm-wavelength
thermal continuum emission of mm1b is similar in intensity and extent to that of mm1a, even if not
as strong, mm1b is not associated with any cm continuum emission down
to a threshold of 1~mJy,  while mm1a is. Moreover, the
molecular spectrum of mm1b seems very different from that of mm1a. Of
all the species we observed, only C$^{34}$S is clearly found in
mm1b. This is not a bias of our observation as other observations confirm our results.
 Class II methanol masers and a mid-IR source are detected only towards mm1a, hence
confirming the different nature of the two cores. The core mm1b could be at an earlier stage of evolution,
when the heating from the central star still has not produced any complex chemistry, nor formed a hypercompact H{\sc ii} region. 

The region around mm2 represents a challenge for the interpretation of
the continuum emission \citep{05358-beuther}, as well as for the
spectral data. Several sub-sources are identified around the main dust
condensation, which are probably caused by the interaction of
molecular outflows with the molecular cloud. Their total mass
corresponds to 5~M$_\odot$.  The molecular emission associated with
the dust core does not show any sign of highly excited lines,
but points out at a warm ($T\sim 60$~K), dense ($n> 6\times
10^6$~cm$^{-3}$) gas with a methanol abundance of $\sim 7\times
10^{-9}$. The $X_{\rm{CH_3OH}}$ of mm2 is more similar to the values
derived in star forming regions of lower mass than to high-mass young
stellar objects. Given the total mass of the region, and the low abundance of methanol, we speculate that
mm2 is a low--intermediate mass protostar. However, the spatial resolution of our data does not allow us to verify whether or not at least part of the emission  from mm2 is indeed not coming from mm2-line.

The coldest and least active core of the region is the
dust condensation mm3. Its mm continuum emission shows that mm3 is a
compact, unresolved source even at the highest resolution reached by
our observations \citep{05358-beuther}. No compact molecular emission is
detected from this core, but all the lines observed toward it
are seen as negative features due to the filtering out of large
structures of the data. The analysis of the spectral energy distribution \citep{05358-beuther}, and of the
gas around the core confirm that mm3 is a dense (n$>10^5$~cm$^{-3}$), cold ($T<40$~K) condensation, surrounded
by a large reservoir of accreting
material \citep[$M\sim 19~M_\odot$,][]{05358-beuther}.
The extended, flat distribution of molecules like CH$_3$OH, C$^{34}$S and H$^{13}$CO$^+$, with no trace of the
structure seen in the dust continuum emission, is reminiscent of low-mass starless cores, where 
C-bearing molecules are freeze-out
onto the grain mantles in a cold and dense environments \citep[e.g.,][]{1999ApJ...523L.165C}. 
Could mm3 be a massive starless core? Massive starless cores can be found in the vicinity of 
active sites of massive star-formation, and should be quiescent displaying no signposts of star formation.
Its properties suggest that mm3 is indeed a massive cold core in a
very early evolutionary stage. However, with the observations we currently have we cannot exclude 
that star formation have already started in this core. In particular, we cannot exclude that
mm3 is the driving source of one of the outflows detected in the region.

\section{Summary}

Our new interferometric data resolve at least four cores in the high-mass protocluster IRAS\,05358+3543.
By analysing  the molecular spectrum of each condensations, we characterised the properties of the
gas surrounding it. Our main results are summarised in the following:

\begin{itemize}
\item the main powerhouse of the region, mm1a, harbours a hot core
with $T\sim 220$~K, and the central heating source has a chemical timescale of
$10^{4.6}$~yr. Our data suggest that mm1a might host a massive
circumstellar disk;
\item  although the properties of 
the mm continuum emission of mm1b are very similar to the one of mm1a, the two
sources differ significantly in the cm and mid-infrared spectrum. This core could be in an earlier stage of star formation than
mm1a, since no molecular emission is detected toward it, with
the only exception of the $5 \to 4$~C$^{34}$S transition;
\item given the low abundance of methanol, mm2 could be a low--intermediate mass protostar; 
\item strong emission is detected in several molecular species to the
north-west of mm2, at a position where no continuum emission is
detected. We suggest that this is caused by the interaction of the
outflows with the ambient molecular cloud;
\item the least active source, mm3, could be a starless massive
core, since it is cold ($T<20$~K), with a large reservoir of accreting
material ($M\sim 19~M_\odot$), but no molecular emission peaks on it.
\end{itemize}

\bibliographystyle{aa}
\bibliography{7977.bib}

\acknowledgements{We would like to thank Jan Martin Winters and Roberto Neri for
their helpful support during the reduction of the PdBI data, and Gabriel Paubert for performing the 
IRAM 30~m observations in service mode. We thank Vincent Minier for providing us with the positions of the velocity channels of the 6.7~GHz methanol maser line. H.B. acknowledges
financial support by the Emmy-Noether-Program of the Deutsche
Forschungsgemeinschaft (DFG, grant BE2578).}

\end{document}